\documentclass[11pt,onecolumn]{article}
\usepackage{amsfonts}
\usepackage{pifont}
\usepackage{amssymb}
\usepackage{amsmath}
\usepackage{multirow}
\usepackage{makecell}
\usepackage{pdflscape}
\usepackage{geometry}
\usepackage{CJK}

\usepackage{caption}
\usepackage{latexsym}
\usepackage{graphicx}
\usepackage{epsfig}

\normalsize \makeatletter
\newenvironment{proof}{{\em Proof.} }{\hfill$\Box$\vspace{0.1in}}

\newtheorem{proposition}{Proposition}[section]

\newtheorem{lemma}{Lemma}[section]

\newtheorem{theorem}{Theorem}[section]

\newcommand{\head}[1]
{\markright{\hbox to 0pt{\vtop to 0pt{\hbox{}\vskip 3mm \hrule
width \textwidth \vss} \hss}{\sc #1}}}

\begin{document}
\title{Online Scheduling with Predicted Order of Processing Times}
\author{Junwen XIAO\thanks{School of Mathematical Science, Zhejiang University, Hangzhou 310027, P. R. China.} \and Tianqi CHEN\thanks{School of Mathematical Science, Zhejiang University, Hangzhou 310027, P. R. China.} \and Zhiyi TAN\thanks{Corresponding author. School of Mathematical Science, Zhejiang University,
Hangzhou 310027, P. R. China. Supported by the National Natural
Science Foundation of China (12571340). {\tt
tanzy@zju.edu.cn}} }
\date{}
\maketitle

\begin{abstract}
In this paper, we investigate the non-clairvoyant online parallel machine scheduling problem with predicted processing time orders, with the objective of minimizing the makespan. A set of jobs is to be processed on $m$ identical parallel machines. The processing time of a job remains unknown until its completion. However, a predicted order of processing times is provided in advance, which may contain errors.  We use $\tau$, the number of inversions between the predicted order and the exact order of processing times, as a metric to measure prediction accuracy. We derive lower bounds of the problem. We also propose two algorithms, POLPT for an arbitrary number of machines and DLPT for three machines, and prove their competitive ratios. Specifically, we establish optimality for $m=2$, for $m=3$ except for $\tau=4$ and $5$, and for $m\geq 4$ with $\tau=2, 3$ or $\tau\geq m^2-m$.
\vskip 2mm\noindent\textbf{Keywords:} Scheduling, Online algorithm, Competitive ratio, Parallel machines, Prediction.
\end{abstract}

\section{Introduction}
Parallel machine scheduling is one of the most important classes of combinatorial optimization problems, for which numerous models and algorithms have been developed. The most fundamental model involves $m$ machines $M_1, M_2, \dots, M_m$ and $n$ jobs $J_1, \dots, J_n$. Each job $J_j$ has a processing time $p_j$, which is its most critical parameter, and must be processed on exactly one of these machines. The scheduler needs to assign each job to a machine and determine its start time to optimize a specific objective function. The most widely studied objective is the minimization of the \textit{makespan}, defined as the maximum completion time among all jobs.

Parallel machine scheduling was first proposed by Graham in 1966~\cite{Graham66}. The model he studied is now referred to as \textit{offline}, where the processing times of all jobs are completely known to the scheduler before making decisions. In 1985, Sleator and Tarjan~\cite{ST85} introduced the concept of \textit{online}, characterized by unknown information at the time of decision-making. The combination of this paradigm with scheduling contexts naturally gave rise to the \textit{online-over-list} scheduling~\cite{FKT89}. In this model, jobs are available at time $0$ and arrive in a given order. For any $j \geq 1$, when assigning job $J_j$, the scheduler only knows the processing times of $J_j$ and the preceding jobs $J_1, \dots, J_{j-1}$. The processing times of subsequent jobs are unknown, and their very existence might not even be known. Furthermore, decisions made for assigned jobs are irrevocable. Since the processing time of a job becomes known upon its arrival, all assignments can be performed immediately at time $0$. There is no benefit in waiting for a job to finish processing before scheduling the next one.

In 1997, Chen and Vestjens proposed another online scheduling model, \textit{online-over-time}, in which each job has a release time \cite{CV97}. The processing time of a job becomes known if and only if its release time has passed, regardless of whether it has started processing. In this setting, the lack of information on processing times is not because previous jobs have not yet been assigned, but simply because the release time has not yet been reached. 
Furthermore, the processing times of all jobs whose release times have passed are known, and there is no sequential dependency as in online over list scheduling. Obviously, it is unwise for the scheduler to assign all jobs at time $0$, and the completion time of a job would be known if it is assigned after its release time.

In addition to these two models, there is also a \textit{non-clairvoyant} online scheduling model~\cite{SWW95}. In this model, all jobs are available at time $0$, but the processing time of each job is revealed only upon its completion.
The scheduler does not need to decide the assignment of all jobs at time $0$. Therefore, making decisions progressively over time is usually preferable. Nevertheless, the model remains harder than online-over-time, as neither the processing time nor the completion time of a job is known when it is scheduled.

Building upon these three types of models, new research directions have gradually emerged. In 1997, Kellerer et al. proposed the \textit{semi-online} scheduling problem \cite{KKST97}. Although the processing times of jobs are still revealed one by one, some additional information is known in advance, such as the total processing time of all jobs. The amount of information available in this model lies between that of offline and online settings, making it possible to design algorithms that outperform pure online algorithms. Semi-online models are usually investigated within the context of the online-over-list model, and the additional information is generally global and accurate. However, there are also other types of information. Seiden et al. proposed another semi-online model, in which jobs are ordered in non-increasing order of processing times \cite{SSW00}. Although the actual processing times are unknown, it is guaranteed that subsequent jobs will not exceed previous ones in processing time, which also facilitates algorithm design. Tan et al. proposed an inexact semi-online model, assuming that the total processing time of all jobs is known to be within a certain interval \cite{TH07}. In other words, the additional information may contain errors, but these errors are bounded and controllable.

With the rapid development and widespread application of technologies such as machine learning, more accurate and efficient prediction methods for unknown information have become available in practice. Consequently, it is possible to obtain predicted processing time of each job, which are more specific than the global information used in previous semi-online scheduling. Of course, predictions may involve inaccuracies, and when leveraging this predicted information to improve algorithm performance, one must also account for such errors. In 2018, Purohit et al. \cite{PSK18} introduced the online scheduling problem with predictions. Bampis et al. \cite{BKLP23} applied the idea of online algorithms with predictions to non-clairvoyant parallel machine scheduling. Each job has a predicted processing time, but its actual processing time is known only after completion. The scheduler schedules jobs progressively over time. In this setting, although the actual processing time remains unknown, the predicted value provides an estimate, which can still improve algorithm performance, especially when the error is small.

In practical scenarios, accurately obtaining or predicting the processing time of a job is often challenging, whereas determining the rank order of processing times is relatively more feasible. Taking hospital surgery scheduling as an illustrative example, the exact duration of a surgical procedure is influenced by multifarious factors, such as individual patient anatomical differences, physiological status, and potential intraoperative complications. These uncertain factors make it difficult for physicians to provide precise predictions of processing times before the surgery. However, physicians can make robust judgments regarding surgical complexity based on the procedure type. For instance, a heart transplantation involves significantly greater technical complexity and resource consumption than an appendectomy. Consequently, it is reasonable to assert that the former necessarily requires a longer duration than the latter. Therefore, leveraging such easy-to-obtain and reliable ordinal information for scheduling holds greater practical value than relying solely on inaccurate numerical predictions.

In this paper, we investigate online scheduling problem with predicted order of processing times. There is an order of jobs, which is a prediction of the non-increasing order of their actual processing times. Predictions may contain errors, but the magnitude of the error is evaluable and known in advance. The scheduler can assign jobs based on the predicted order, while the actual processing times are only revealed upon job completion. Job assignments can be made progressively over time. Of course, the predicted order could also be completely correct, in which case the problem reduces to a semi-online non-clairvoyant scheduling problem where jobs are known in non-increasing order of processing times.

A similar problem is the ordinal online scheduling, which was proposed by Liu et al. in 1996 \cite{LSV96}. In this model, the exact order of jobs by their processing times is also known, but the assignment of all jobs must be made at time $0$ simultaneously.
Since the processing times of all jobs are completely unknown at the time of assignment, it is irrelevant when they become known.

{\bf Previous Work: } For offline and various online scheduling problems, the performance of algorithms is typically evaluated using the \textit{worst-case ratio} and the \textit{competitive ratio}, respectively. For an online problem, if no algorithm can achieve a competitive ratio smaller than $\rho$, then $\rho$ is termed the \textit{lower bound} of the problem, and an algorithm achieving this ratio is called \textit{optimal}. For a comprehensive overview on online scheduling problems and algorithms, readers are referred to the survey \cite{PST04, TZ13}.

Graham introduced the \textit{List Scheduling} (LS) algorithm in his seminal paper, proving its worst-case ratio to be $2 - \frac{1}{m}$ \cite{Graham66}. The LS algorithm assigns jobs in their given order to the machine that can start processing the job earliest. Subsequently, Graham proposed an improved algorithm, \textit{Longest Processing Time first} (LPT), achieving a worst-case ratio of $\frac{4m-1}{3m}$ \cite{Graham69}. The LPT algorithm operates by first sorting jobs in non-increasing order of processing times and then assigning them according to the LS rule. Although the formal framework of online scheduling was not established at the time, the LS algorithm is inherently online and has been proven optimal for $m=2, 3$ in the online-over-list model \cite{FKT89}. Since then, lower bounds and competitive ratios have been continuously refined. Currently, the best known lower bound is $1.85358$ \cite{GRTW00}, while the best known competitive ratio is $1.9201$ \cite{FW00}. If jobs are presented in non-increasing order of processing times, there exists an algorithm with a competitive ratio of at most $\frac{5}{4}$ \cite{CKK12}. Specifically, for $m=2$ and $3$, the competitive ratios of the optimal algorithms are $\frac{7}{6}$ and $\frac{1+\sqrt{37}}{6}$, respectively \cite{SSW00, CKK12}.

For the online-over-time model, a lower bound of $1.3473$ has been established, and the competitive ratio of the LPT algorithm is $\frac{3}{2}$ \cite{CV97}. Specifically, for $m=2$, the competitive ratios of the optimal algorithm is $\frac{5-\sqrt{5}}{2}$ \cite{NS01}. A slightly improved algorithm for $m\geq 3$ was recently proposed in \cite{WYZ25}.
For the non-clairvoyant online model, the competitive ratio of the LS algorithm is $2 - \frac{1}{m}$, which is optimal for any number of machines \cite{SWW95}.

For online scheduling with predictions, Bampis et al. introduced $\varphi$ as an indicator to measure the accuracy of prediction, defined as the maximum over all jobs of the larger value between the ratio of predicted to actual processing time and the ratio of actual to predicted processing time. They then expressed the competitive ratio and the lower bound as functions of $\varphi$ \cite{BKLP23}. They modified the LPT algorithm by sequencing jobs in non-increasing order of predicted processing times. With the improvement proposed in \cite{CT26}, the competitive ratio of the algorithm is given by
\begin{equation*}
	\begin{cases}
		1+\frac{(m-1)\varphi^2}{3m}, & 1\leq \varphi\leq \sqrt{\frac{3m}{m+2}},\\
		\frac{1+\varphi^2}{2}, & \sqrt{\frac{3m}{m+2}}\leq \varphi\leq \sqrt{\frac{3m-2}{m}},\\
		2-\frac{1}{m}, & \varphi\geq \sqrt{\frac{3m-2}{m}},
	\end{cases}
\end{equation*}
while the lower bound of the problem is:
\begin{equation*}
	\begin{cases}
		\frac{1+\varphi^2}{2}, & \varphi\leq \sqrt{2},\\
		2-\frac{1}{\varphi^2}, & \sqrt{2}<\varphi< \sqrt{m},\\
		2-\frac{1}{m}, & \varphi\geq \sqrt{m}.
	\end{cases}
\end{equation*}
Thus, the algorithm is optimal when $\varphi\geq \sqrt{m}$. For $m=2,3$, the competitive ratios were further improved, and the algorithms were proven to be optimal when $\varphi\geq \sqrt[4]{m}$.

For ordinal online scheduling on an arbitrary number of machines, the best known lower bound and competitive ratio are $\frac{3}{2}$ and $\frac{5}{3}$, respectively \cite{LSV96}. For a small fixed number of machines, both the lower bounds and the algorithms can be improved \cite{LSV96, Epstein23}. Specifically, optimal algorithms for $m=2$ and $3$ achieve competitive ratios of $\frac{4}{3}$ and $\frac{7}{5}$, respectively \cite{LSV96}.

Recently, several papers have addressed scheduling problems where a certain order of jobs is predictable. 
Fu et al. \cite{FZSQSZ24} investigated scheduling problems motivated by LLM inference and serving with the objectives of minimizing average latency and makespan. Each request has an unknown output length, measured by the number of generated tokens, which determines its processing time. The latency of a request is defined as its flowtime divided by its output length, where flowtime is the difference between its completion time and its release time. Although the exact output lengths are unknown at the time of scheduling, their relative order can be predicted. The prediction error is quantified using the Kendall tau coefficient \cite{Kendall45} between the predicted and true orders, which is a linear function of the number of inversions. They propose several heuristics, but provide no theoretical performance guarantees. Lindermayr and Megow \cite{LM25} studied scheduling on single and parallel machines with the objective of minimizing total weighted completion time. Since jobs are characterized by both processing time and weight, the predicted order is based on a composite metric of the two, resulting in a more complex definition of prediction error. They design algorithms and prove their competitive ratios, all of which require the optimal value as a parameter.

\begin{figure}
	\includegraphics[width=160mm]{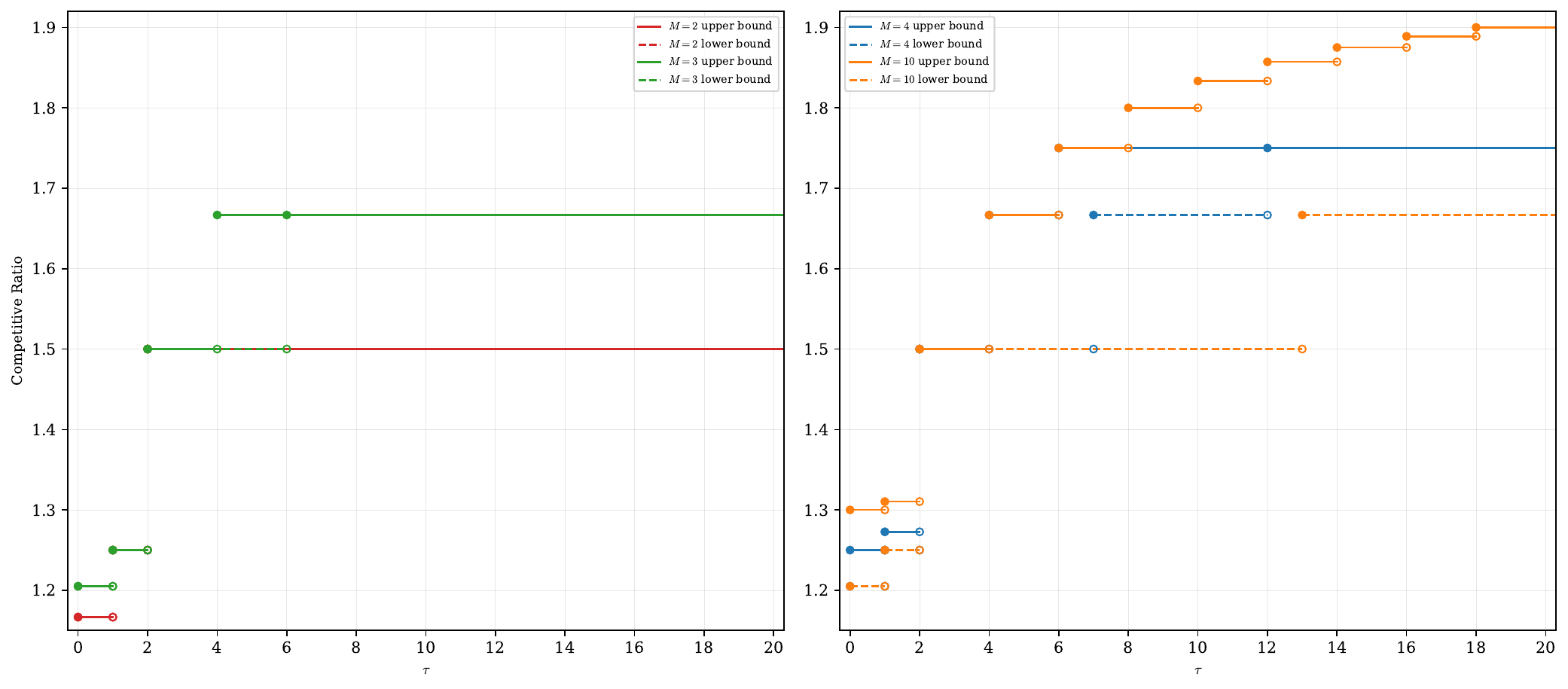}
	\centering
	\caption{Lower bounds of the problem and competitive ratios of the algorithms for $m=2,3,4,10$. }\label{fig1}
\end{figure}

{\bf Our Work: }
In this paper, we study the non-clairvoyant online scheduling problem on $m$ machines with predictions regarding the order of jobs by their processing times. We use the number of inversions $\tau$ between the actual processing time order and the predicted processing time order as an indicator to measure the accuracy of prediction. This metric has also been employed in the literature \cite{FKMV16}. We design an algorithm based on the predicted order and prove that its competitive ratio, as a function of $\tau$, is
\begin{equation}\label{t70}
	r_{\tau,m}^P=\begin{cases}
		\frac{4m-1}{3m} & \text{if } \tau=0, \\
		\max\left\{\frac{5}{4},\frac{4m-2}{3m-1}\right\} & \text{if } \tau= 1, \\
		2 - \frac{1}{\left\lfloor \frac{\tau}{2} \right\rfloor + 1} & \text{if } 2\leq \tau \leq 2m-2, \\
		2 - \frac{1}{m} & \text{if } \tau > 2m-2.  
	\end{cases}	
\end{equation} 
We also establish that the lower bound of the problem is
\begin{equation*}
	\begin{cases}
		\frac{7}{6} & \text{if } m=2, \tau=0,  \\
		\frac{13+\sqrt{15}}{14}\approx 1.205 & \text{if } m\geq 3, \tau=0, \\
		\frac{5}{4} & \text{if } m\geq 2, \tau= 1,  \\
		2 - \frac{1}{m - t} & \text{if } m\geq 2, m^2-(m+1)t-m\leq \tau\leq m^2-(m+1)t, 1\leq t\leq m-2, \\
		2 - \frac{1}{m} & \text{if } m\geq 2, \tau\geq m^2-m.  
		  
	\end{cases}	
\end{equation*}  
For the case where predictions are perfectly accurate and $m=3$, we propose an algorithm with a competitive ratio of $\frac{13+\sqrt{15}}{14}$. Thus, optimal algorithms are obtained for any $\tau \geq 0$ when $m=2$, for $0\leq \tau \leq 3$ and $\tau\geq 6$ when $m=3$, and for $2\leq \tau \leq 3$ and $\tau \geq m^2-m$ when $m\geq 4$. Figure \ref{fig1} illustrates the best competitive ratios of the proposed algorithm and the best lower bounds obtained for various values of $m$ and $\tau$.

{\bf Structure of the paper: } In Section 2, we formally define the problem and present some preliminary results. The algorithm for a general number of machines and an improved algorithm for $m=3$ are analyzed in Sections 3 and 4, respectively. In Section 5, we derive lower bounds for the problem. Finally, conclusions are given in Section 6.

\section{Preliminaries}

An instance consists of a set of jobs $\mathcal{J}=\{J_1, \dots, J_n\}$ and a set of machines $\mathcal{M}=\{M_1, \dots, M_m\}$. Jobs are indexed in non-increasing order of their processing times, i.e., $p_1 \geq p_2 \geq \dots \geq p_n$. In this paper, we use a permutation $\pi=(j_1, j_2, \dots, j_n)$ of $(1,2,\dots,n)$ to represent an order of jobs, where job $J_{j_i}$ is in the $i$-th position. Write $\pi(i)=j_i$ and $\pi^{-1}(j_i)=i$.  We use the terms permutation and order interchangeably. The permutation $(1, 2, \dots, n)$ represents the non-increasing order of the jobs' actual processing times. 

Although the actual processing times are not known in advance, we can predict their relative order by some prediction method. However, the predicted order may not coincide with the true non-increasing order $(1,2,\dots,n)$. Thus, a metric is required to quantify the difference between two orders, reflecting the prediction accuracy. In this paper, we define this metric by the number of inversions in the predicted order.

For any $j,l\in \{1,2,\dots,n\}$, if $j < l$ but $\pi(j)>\pi(l)$, the pair $(j, l)$ is called an {\it inversion} of $\pi$. The total number of inversions in the permutation $\pi$ is called the {\it inversion number}, denoted by $\tau(\pi)$, which is simplified as $\tau$ when no confusion arises.  
The following lemma presents some properties of the inversion number.

\begin{lemma}\label{l0}
(i)	$\tau \geq n - \pi(n)$, and if $\tau= n - \pi(n)$, then 
\begin{equation}\label{e9}
\pi(j)=\left\{ 
\begin{array}{ll}
	j & 1 \leq j \leq n - \tau - 1, \\
	j + 1 & n - \tau \leq j \leq n - 1, \\
	n - \tau & j = n.  
\end{array} 
\right.	
\end{equation}
(ii) $\tau \geq \pi^{-1}(1)-1$.
\end{lemma}
\begin{proof}
(i)	If $\pi(n)=n$, the conclusion holds trivially. Otherwise, there exist $n - \pi(n)$ inversions $(\pi^{-1}(j),n)$ in $\pi$, where $j=\pi(n)+1,\ldots,n$. Therefore, $\tau \geq n - \pi(n)$. If $\tau= n - \pi(n)$, there can be no other inversions in $\pi$. Consequently, for any $j<l$ and $l\neq n$, $\pi(j)<\pi(l)$. It follows that $\pi(j)=j$ for any $1\leq j<\pi(n) $, and $\pi(j)=j+1$ for any $\pi(n)\leq j<n $.

(ii) For any $1\leq j<\pi^{-1}(1)$, $(j,\pi^{-1}(1))$ forms an inversion, which implies $\tau \geq \pi^{-1}(1)-1$.
\end{proof}


However, it is worth noting that a particularity arises when using a permutation to denote the order of non-increasing job processing times. If $j < l$ and $p_{\pi(j)} = p_{\pi(l)}$, predicting $\pi(j) > \pi(l)$ should not be considered an error in practice. Therefore, we define a pair $(j, l)$ with $j < l$ and $p_{\pi(j)}<p_{\pi(l)}$ as a \textit{true inversion}. The total number of true inversions in permutation $\pi$ is denoted by $\overline{\tau}(\pi)$. The following lemma demonstrates how the number of inversions and true inversions in a permutation changes after swapping a pair that is an inversion but not a true inversion.

\begin{lemma}\label{ln1}
	Let $(j, l)$ be an inversion of a permutation $\pi$ with $p_{\pi(j)} = p_{\pi(l)}$. Let $\pi'$ be the permutation obtained by transposing the entries at positions $j$ and $l$, i.e.,
	\[
	\pi'(i) = 
	\begin{cases} 
		\pi(l), &  i = j, \\ 
		\pi(j), &  i = l, \\ 
		\pi(i), &  i \neq j, l.
	\end{cases}
	\]
	Then $\tau(\pi') < \tau(\pi)$ and $\overline{\tau}(\pi') = \overline{\tau}(\pi)$.	
\end{lemma}
\begin{proof}
	For any $i < k$ and neither $i$ nor $k$ is equal to $j$ or $l$,  $\pi(i) = \pi'(i)$ and $\pi(k) = \pi'(k)$. Hence, $(i, k)$ is an inversion (resp. true inversion) in $\pi$ if and only if it is an inversion (resp. true inversion) in $\pi'$. Note that $\pi(j) = \pi'(l)$, $\pi(l)=\pi'(j)$. For any $1 \leq i \leq j-1<j<l$, $(i, j)$ is an inversion (resp. true inversion) in $\pi$ if and only if $(i, l)$ is an inversion (resp. true inversion) in $\pi'$, and $(i, l)$ is an inversion (resp. true inversion) in $\pi$ if and only if $(i, j)$ is an inversion (resp. true inversion) in $\pi'$. For any $j<l<l+1 \leq k \leq n$,  $(j, k)$ is an inversion (resp. true inversion) in $\pi$ if and only if $(l, k)$ is an inversion (resp. true inversion) in $\pi'$, and $(l, k)$ is an inversion (resp. true inversion) in $\pi$ if and only if $(j, k)$ is an inversion (resp. true inversion) in $\pi'$. 
	
	Next, consider the case where $j < i < l$. 
	If $\pi(i) > \pi(j) > \pi(l)$, then $(j, i)$ is not an inversion in either $\pi$ or $\pi'$, but $(i, l)$ is an inversion in both $\pi$ and $\pi'$. 
	If $\pi(j) > \pi(l) > \pi(i)$, then $(j, i)$ is an inversion in both $\pi$ and $\pi'$, but $(i, l)$ is not an inversion in either $\pi$ or $\pi'$. 
	If $\pi(j) > \pi(i) > \pi(l)$, then both $(j, i)$ and $(i, l)$ are inversions in $\pi$, but neither of them is an inversion in $\pi'$. 
	If $p_{\pi(i)} = p_{\pi(j)} = p_{\pi(l)}$, then neither $(j, i)$ nor $(i, l)$ is a true inversion in either $\pi$ or $\pi'$. 
	If $p_{\pi(i)} < p_{\pi(j)} = p_{\pi(l)}$, then $(j, i)$ is not a true inversion in either $\pi$ or $\pi'$, but $(i, l)$ is a true inversion in both $\pi$ and $\pi'$. 
	If $p_{\pi(j)} = p_{\pi(l)} < p_{\pi(i)}$, then $(j, i)$ is a true inversion in both $\pi$ and $\pi'$, but $(i, l)$ is not a true inversion in either $\pi$ or $\pi'$.
	
	Finally, $(j, l)$ is an inversion in $\pi$ but not in $\pi'$, and it is not a true inversion in either $\pi$ or $\pi'$. 
	Combining the above cases, we conclude that $\tau(\pi') < \tau(\pi)$ and $\overline{\tau}(\pi') = \overline{\tau}(\pi)$.
\end{proof}

An algorithm is called {\it lazy} if it only relies on the permutation of jobs, but does not utilize any information about the inversion number or true inversion number of the permutation. The following lemma shows that for lazy algorithms, there is no essential difference between the inversion number and the true inversion number in terms of the competitive ratio. 

\begin{lemma}\label{ln2}
	If for any $\tau\geq 0 $ and any instance with inversion number at most  $\tau$, 	$\frac{C^A(I)}{C^*(I)}\leq r_{\tau,m}^A$, then for any instance with true inversion number at most  $\tau$, $\frac{C^A(I)}{C^*(I)}\leq r_{\tau,m}^A$.
\end{lemma}
\begin{proof}
Consider any instance $I$ with a job prediction order given by a permutation $\pi$ satisfying $\overline{\tau}(\pi) \leq \tau$. If there exist $j < l$ such that $\pi(j) > \pi(l)$ and $p_{\pi(j)} = p_{\pi(l)}$, we can construct a new permutation $\pi'$ by transposing the entries at positions $j$ and $l$. By Lemma \ref{ln1}, $\tau(\pi') < \tau(\pi)$ and $\overline{\tau}(\pi') = \overline{\tau}(\pi)$. We repeat this process until for all $j < l$ and $\pi'(j) > \pi'(l)$, $p_{\pi'(j)} < p_{\pi'(l)}$. Consequently, every inversion in $\pi'$ is also a true inversion, which implies $\tau(\pi') = \overline{\tau}(\pi') = \overline{\tau}(\pi) \leq \tau$.
	
	Now, apply Algorithm A to a modified instance $I'$, which shares the same job set with $I$ but uses $\pi'$ as its prediction order. By our assumption, $\frac{C^A(I')}{C^*(I')} \leq r_{\tau,m}^A$.
	Since the processing times of the jobs remain unchanged, and the processing time sequences induced by $\pi$ and $\pi'$ are identical, we have $C^A(I) = C^A(I')$ and $C^*(I) = C^*(I')$. Therefore,
	$\frac{C^A(I)}{C^*(I)} = \frac{C^A(I')}{C^*(I')}\leq r_{\tau,m}^A$, which completes the proof.	
\end{proof}

The lemma implies that to establish the competitive ratio of an algorithm under a given upper bound on the true inversion number, it suffices to analyze its performance with respect to the inversion number. Conversely, if it is proven that no algorithm can achieve a competitive ratio below a certain value for a specific upper bound on the true inversion number, then no such algorithm exists for the inversion number under the same bound.

Given a subset $\mathcal{J}_0\subseteq \mathcal{J}$, the total processing time of the jobs in $\mathcal{J}_0$ is denoted by $P(\mathcal{J}_0)$. We abbreviate $P(\mathcal{J})$ to $P$. For any schedule $\sigma^A$, denote by ${\cal J}_i^A$ the set of jobs scheduled on $M_i$ in $\sigma^A$ and $L_i^A=P({\cal J}_i^A)$ the {\it load} of $M_i$ in $\sigma^A$, $i=1\dots,m$. Let $C^A_j$ be the completion time of $J_j$ in $\sigma^A$, $j=1\dots,n$.
The makespan of $\sigma^A$ is denoted $C^A$, and $C^A=\max\limits_{i=1,\dots,m}L_i^A=\max\limits_{j=1,\dots,n}C^A_j$. Let ${\cal M}_l^A = \{ M_k \mid |{\cal J}_i^A| = l \}$ be the set of machines that process $l$ jobs in $\sigma^A$ and $\mu_l^A=|{\cal M}_l^A|$ be the number of machines in ${\cal M}_l^A$, $l\geq 1$. Obviously,  if $\mu_1^A+\mu_2^A=m$, then 
\begin{equation}\label{t11}
	n=2m-\mu_1^A.	
\end{equation} 

Let $\sigma^*(I)$ be the optimal schedule of $I$. We will drop $I$ if no confusion arises. Below is a property of an optimal schedule, which can be proven using the standard job interchange argument. If an optimal schedule fails to satisfy this property, we can obtain a new schedule that is both optimal and satisfies the required property by performing a series of changes to the machines assigned to jobs, or swaps of machines between two jobs.

\begin{lemma}\label{l5}
	If $n \geq m + 1$, then there exists an optimal schedule $\sigma^*$ such that
	
	(i) ${\cal J}_i^* \neq \emptyset$ for any $1 \leq i \leq m$. 
	
	(ii) If no two jobs of $ \{J_1, J_2, \dots, J_{j}\} $ are scheduled on the same machine, then $J_i\in {\cal J}_i^*$, $i=1,\ldots,j$. 
	
	(iii) If $\mu_1^* > 0$, then ${\cal J}_i^*=\{J_i\}$, $i=1,\dots, \mu_1^*$. 
	
	(iv) If $1\leq |{\cal J}_i^*|\leq 2$ for any $i$, then 
	$${\cal J}_i^*=\{J_i\}, i=1,\dots, \mu_1^*, {\cal J}_i^*=\{J_i, J_{2m-i+1}\}, i=\mu_1^*+1,\dots, m. $$
\end{lemma}

\section{The Algorithm POLPT}

In this section, we present the algorithm for an arbitrary number of machines. We start by analyzing the most intuitive algorithm, POLPT (Predicted Order LPT), which assigns jobs to idle machines sequentially according to the predicted order.

\textbf{Algorithm POLPT}

Whenever a machine becomes idle, it always assigns to that machine the unassigned job with the smallest position in the prediction order.

Obviously, POLPT is a lazy algorithm. Below we prove that the competitive ratio of POLPT is $r_{\tau,m}^P$, as given by \eqref{t70}, where $\tau$ is an upper bound on the inversion number or the true inversion number of the predicted order, and $m$ is the number of machines. Note that $r_{\tau,m}^P$ is non-decreasing with respect to both $\tau$ and $m$.

Denote by $\sigma^P(I)$ the schedule of $I$ produced by algorithm POLPT. Let $I_j$ be the instance with the same number of machines as $I$ and the job set $\{J_{\pi(1)}, J_{\pi(2)}, \dots, J_{\pi(j)}\}$  consisting of the $j$ jobs with the smallest  positions in the predicted order of $I$. Based on the rule of POLPT, it is not difficult to see that the assignment of the jobs of $I_j$ in $\sigma^P(I_j)$ is the same as that in $\sigma^P(I)$.


If $J_{\pi(j)}$, $j < n$, determines the makespan, then $C^P(I_j) = C^P(I)$. Since $C^*(I_j) \leq C^*(I)$, we have $\frac{C^P(I)}{C^*(I)} \leq \frac{C^P(I_j)}{C^*(I_j)}$. Hence, to prove the competitive ratio of the POLPT, it is sufficient to consider the instances that $J_{\pi(n)}$ is the only job that determines the makespan. 

Denote by $s_n$ the start time of $J_{\pi(n)}$ in $\sigma^P$. By the rule of the POLPT, all the machines would not be idle before time $s_n$. Hence
\begin{equation}\label{e1}
C^{P} = s_n + p_{\pi(n)} \leq \frac{P - p_{\pi(n)}}{m} + p_{\pi(n)} = \frac{P}{m} + \frac{m-1}{m} p_{\pi(n)}	
\end{equation}
and
\begin{equation}\label{e2}
	C^{*} \geq \frac{P}{m}\geq \frac{ms_n + p_{\pi(n)}}{m}=s_n +\frac{1}{m}p_{\pi(n)}.	
\end{equation}
By (\ref{e1}), (\ref{e2}) and $C^*\geq \max\limits_{j=1,\dots,n}p_j$, we have the following lemma. 
\begin{lemma}\label{l2}
	(i) $\frac{C^P}{C^*} \leq 2 - \frac{1}{m}$.
	
	(ii) If $C^* \geq 3p_{\pi(n)}$, then $\frac{C^P}{C^*} \leq \frac{4m-1}{3m}$.
\end{lemma}

Since $\frac{4m-1}{3m}\leq r_{\tau,m}^P$ for any $\tau\geq 0$ and $m\geq 2$, we assume that $C^*<3p_{\pi(n)}$ in the following. As there must be a machine that processes three of the $2m+1$ jobs with the largest processing times in $\sigma^*$, we have $C^* \geq 3p_{2m+1}$. 
Thus $\pi(n) \leq 2m$ and $n \leq 2m + \tau$ by Lemma \ref{l0}(i).

\begin{lemma}\label{l6}
	If $n \leq m$, then $\sigma^P$ is the optimal schedule.
\end{lemma}
\begin{proof}
Note that the first $\min\{m,n\}$ jobs are assigned by POLPT to separate machines. If $n \leq m$, $|{\cal J}_i^P|\leq 1$ for any $1\leq i\leq m$ and thus  $C^{P} = p_{1}\leq C^*$. 
\end{proof}

Therefore, to prove the competitive ratio of POLPT, we only need to consider the case of $n> m+1$. Then $|{\cal J}_i^P|\geq 1$ for any $i$  and $s_n>0$. We assume w.l.o.g. that $J_{\pi(i)} \in {\cal J}_i^P$ for any $1\leq i\leq m$. Equivalently, if $J_j$ is the first job assigned to a machine, then that machine must be $M_{\pi^{-1}(j)}$.  

Next, we present several properties of $\sigma^P$.

\begin{lemma}\label{l7}
In  $\sigma^P$, no two jobs with processing times greater than $s_n$ are scheduled on the same machine.	
\end{lemma}
\begin{proof}
Assume for the sake of contradiction that jobs $J_{\pi(j)}$ and $J_{\pi(l)}$ are scheduled on the same machine, where $j<l$ and $\min\{p_{\pi(j)}, p_{\pi(l)}\}>s_n$. Then $C_{\pi(j)}^P \geq p_{\pi(j)} > s_n$, which implies that the start time of $J_{\pi(l)}$ exceeds $s_n$. Thus, $l<n$, and $J_{\pi(l)}$ should be assigned to the machine processing $J_{\pi(n)}$. This is a contradiction.
\end{proof}	

\begin{lemma}\label{l3}
	Suppose that $\{\pi(1), \pi(2), \dots, \pi(m)\} = \{j_1, j_2, \ldots, j_m\}$ with $j_1 <j_2< \ldots< j_m$. For any $1\leq l\leq m-1$, each of $M_{\pi^{-1}(j_1)}, \ldots, M_{\pi^{-1}(j_{m-l})}$ processes exactly one job after the assignment of $J_{\pi(m+l)}$, and $C_{\pi(m+l)}^P \leq p_{j_{m-l+1}} +p_{\pi(m+l)} $. The equality holds when $J_{\pi(m+l)} \in {\cal J}_{\pi^{-1}(j_{m-l+1})}^P$.
\end{lemma}
\begin{proof}
	We prove the lemma by induction on $l$. Note that $J_{j_i}$ is the first job scheduled on $M_{\pi^{-1}(j_{i})}$ for any $i$.  Since 
	$p_{j_m}= \min\limits_{i=1,\dots,m}p_{j_i}$,
	$J_{\pi(m+1)}$ is assigned to $M_{\pi^{-1}(j_{m})}$, and the result holds for $l = 1$. Suppose that after the assignment of $J_{\pi(m+1)}, \cdots, J_{\pi(m+l-1)}$, each of $M_{\pi^{-1}(j_1)}, \ldots, M_{\pi^{-1}(j_{m-l+1})}$ processes exactly one job. 
	Since $p_{j_{m-l+1}}= \min\limits_{i=1,\dots,m-l+1}p_{j_i}$, even if $J_{\pi(m+l)}$ is not processed on one of $M_{\pi^{-1}(j_{m-l+2})}, \dots, M_{\pi^{-1}(j_{m})}$, POLPT prefers to assign it on $M_{\pi^{-1}(j_{m-l+1})}$ rather than on any of $M_{\pi^{-1}(j_1)}, \ldots, M_{\pi^{-1}(j_{m-l})}$. If it is processed on $M_{\pi^{-1}(j_{m-l+1})}$, its completion time is exactly $p_{j_{m-l+1}}+p_{\pi(m+l)}  $.  If $J_{\pi(m+l)}$ is not processed on $M_{\pi^{-1}(j_{m-l+1})}$, it is only because its completion time on one of $M_{\pi^{-1}(j_{m-l+2})}, \dots, M_{\pi^{-1}(j_{m})}$ is no more than its completion time on $M_{\pi^{-1}(j_{m-l+1})}$. Therefore, after the assignment of $J_{\pi(m+l)}$, each of $M_{\pi^{-1}(j_1)}, \ldots, M_{\pi^{-1}(j_{m-l})}$ processes exactly one job, and $C_{\pi(m+l)}^P \leq p_{j_{m-l+1}} +p_{\pi(m+l)} $. The result also holds for $l$.  
\end{proof}

The following lemma demonstrates that any instance satisfying a specific property can be reduced to one with fewer machines without compromising the worst-case performance of the algorithm. Given that $r_{\tau,m}^P$ is non-increasing with respect to $m$ and the algorithm is trivially optimal for $m=1$, the proof for these instances follows by induction on $m$. Consequently, it suffices to consider only the instances that do not satisfy this property in the subsequent analysis.
\begin{lemma}\label{l4}
	If ${\cal J}_1^*=\{J_{1}\}$ and ${\cal J}_{\pi^{-1}(1)}^P \subseteq \{J_{1},J_{\pi(n)}\}$, then $\frac{C^P(I)}{C^*(I)} \leq \frac{C^P(I')}{C^*(I')}$, 
	where $I'$ is the instance with job set ${\cal J} \setminus \{J_{1}\}$ and $m-1$ machines.
\end{lemma}
\begin{proof}
	Obviously, $C^*(I') \leq C^*(I)$. Since $\mathcal{J}_{\pi^{-1}(1)}^P \subseteq \{J_{1}, J_{\pi(n)}\}$, $J_{1}$ is the first job processed on $M_{\pi^{-1}(1)}$, and $J_{\pi(n)}$ is the only other possible job on this machine. Therefore, excluding the last job $J_{\pi(n)}$, the assignment of the remaining jobs in $\sigma^P(I')$ is identical to that in $\sigma^P(I)$. In $\sigma^P(I)$, $J_{\pi(n)}$ is assigned to the one with the smaller starting time between $M_{\pi^{-1}(1)}$ and the machine processing it in $\sigma^P(I')$. Consequently, $C^P(I) \leq C^P(I')$, which implies $\frac{C^P(I)}{C^*(I)} \leq \frac{C^P(I')}{C^*(I')}$.
\end{proof}	

With these preparations, we prove the upper bound on the competitive ratio of the algorithm by cases in the following lemmas.

\begin{lemma}\label{m1}
If $\pi(n) = n$, then $\frac{C^P}{C^*} \leq \frac{4m-1}{3m}$.
\end{lemma}
\begin{proof}
Since $\pi(n) = n$, it suffices to consider $C^* < 3p_n$ by Lemma \ref{l2}(ii). Then $|{\cal J}_i^*| \leq 2$ for any $1\leq i\leq m$, and thus $n\leq 2m$. We first claim that $|{\cal J}_i^P| \leq 2$ for any $1\leq i\leq m$. In fact, if there exists $i$ such that $|{\cal J}_i^P| \geq 3$, then $m>\mu_1^P +\mu_2^P$. Moreover, ${\cal M}_1^P \neq \emptyset$ as $n \leq 2m$. Suppose that $\{ J_{\pi(i)}, J_{\pi(j)}, J_{\pi(l)} \}\subseteq {\cal J}_i^P$ and $i< j <l$.  For any $M_k \in {\cal M}_1^P$, $p_{\pi(k)} \geq p_{\pi(i)} + p_{\pi(j)} \geq 2p_n$ by the rule of POLPT. Hence, $J_{\pi(k)}$ is scheduled separately in $\sigma^*$ as $C^* < 3p_n$. Therefore, $\mu_1^P \leq \mu_1^*$ and thus 
$$n\geq 3(m-\mu_1^P-\mu_2^P)+ 2\mu_2^P +\mu_1^P > 2m -\mu_1^P \geq   2m - \mu_1^*=n$$
by \eqref{t11}, a contradiction. 

Suppose that $J_{\pi(n)}$ is scheduled together with $J_k$ on $M_{\pi^{-1}(k)}$. 
For any $1 \leq j \leq k-1$, if $J_j$ is scheduled on the same machine with another job $J_l$, then the load of that machine is $p_j+p_l\geq p_k+p_n=p_k+p_{\pi(n)}=C^P$. This contradicts the assumption that $J_{\pi(n)}$ is the only job determining the makespan. Therefore, $\mathcal{J}_{\pi^{-1}(j)}^P = \{ J_j \}$ for any $1 \leq j \leq k-1$. For any $M_i \in \mathcal{M}\setminus \{ M_{\pi^{-1}(1)}, \cdots, M_{\pi^{-1}(k)} \}$,  $|\mathcal{J}_i^P| = 1$ only if  $p_{\pi(i)} =p_k$. Thus $p_k=p_{k+1}=\dots=p_{\mu_1^P+1}$. Since $n=2m - \mu_1^*=2m - \mu_1^P$ by \eqref{t11}, we have 
$$C^*\geq p_{\mu_1^*+1} + p_n = p_{\mu_1^P+1} + p_n=p_k+p_n= C^P$$ 
by Lemma \ref{l5}(iv).
\end{proof}

\begin{lemma}\label{m2}
	If $\tau =1$, then $\frac{C^P}{C^*} \leq \max\{\frac{5}{4},\frac{4m-2}{3m-1}\}$.
\end{lemma}
\begin{proof}
By Lemmas \ref{l0}(i) and \ref{m1}, we only need to consider the case $\pi(n)=n - 1$. Furthermore, by Lemma \ref{l2}(ii), it suffices to assume $C^* < 3p_{n-1}$. Combining this with Lemma \ref{l6} and $C^* \geq 3p_{2m+1}$, we may assume that $m+1 \leq n \leq 2m + 1$.
	
Note that $\pi(i)=i$ for any $1\leq i\leq m$ by Lemma \ref{l0}(i). Thus $J_i\in {\cal J}_i^P$ for any $1\leq i\leq m$. If $n\leq 2m$, then ${\cal J}_1^P \subseteq \{J_{1},J_{\pi(n)}\}$ by Lemma \ref{l3}. If $\mu_1^* > 0$, then $\frac{C^P}{C^*} \leq r_{\tau, m-1}^P\leq r_{\tau, m}^P$ by Lemmas \ref{l5}(iii) and \ref{l4}. Otherwise, $|J_i^*| \geq  2$ for any $i$. Then $n = 2m$ and $C^* \geq \max \{ p_1 + p_n, p_2 + p_{n-1} \}$ by Lemma \ref{l5}(iv). By Lemma \ref{l3}, the load of $M_1$ and $M_2$ before the assignment of $J_{\pi(n)}$ is $p_1$ and at most $p_2 + p_{\pi(n-1)}$, respectively. Thus 
\begin{align*}
	C^P &\leq \min \{ p_1 + p_{\pi(n)}, p_2 + p_{\pi(n-1)} + p_{\pi(n)} \}=\min \{ p_1 + p_{n-1}, p_2 + p_n + p_{n-1} \} \\
	& \leq \frac{1}{2}(p_1 + p_{n-1}+p_2 + p_{n-1}+ p_n)\leq \frac{1}{2}(p_1 + p_{n}) +  \frac{3}{4}(p_2 + p_{n-1}) \\
	& \leq \frac{5}{4} \max\{p_1+p_n, p_2+p_{n-1}\} \leq \frac{5}{4}C^*.
\end{align*}

If $n = 2m+1$, there exists $k$ such that $|{\cal J}_k^*| =3$. Hence, $C^* \geq 2p_{n-1} + p_n$. Since $C^* < 3p_{n-1}$, any machine that processes $3$ jobs must process $J_n$. Thus $|{\cal J}_i^*| =2$ for any $i\neq k$ and $C^* \geq p_1 + p_{n-1}$. By Lemmas \ref{l3} and \ref{l0}(i), $M_1$ processes only one job before the assignment of $J_{\pi(2m)}$ and $\pi(2m)=n$. Hence, $s_n\leq p_1+p_{\pi(2m)}=p_1+p_{n}$. Combining (\ref{e2}) with the above discussions, we have
\begin{align*}
	C^P &= s_n+p_{\pi(2m+1)}=s_n+p_{n-1}\leq 
	\frac{1}{3m-1}\left(  2m(s_n + p_{n-1}) + (m-1)(p_1 + p_n + p_{n-1}) \right) \\
	&=\frac{1}{3m-1}\left( 2m\left( s_n + \frac{1}{m}p_{n-1}\right)  + (m-1)(p_1 + p_{n-1}) + (m-1)(2p_{n-1} + p_n) \right) \\ 
	&\leq \frac{1}{3m-1}\big( 2mC^* + (m-1)C^* + (m-1)C^* \big) = \frac{4m-2}{3m-1}C^*.
\end{align*}
\end{proof}	

\begin{lemma}\label{m3}
	If $2\leq \tau \leq 2m-2$, then $\frac{C^P}{C^*} \leq 2 - \frac{1}{\left\lfloor \frac{\tau}{2} \right\rfloor + 1}$.
\end{lemma}
\begin{proof}
If $C^*\geq 2p_{\pi(n)}$, then
$$C^P = s_n +p_{\pi(n)} \leq \frac{3}{2} C^*\leq  \left( 2 - \frac{1}{\left\lfloor \frac{\tau}{2} \right\rfloor + 1}\right) C^*$$
by (\ref{e2}). Hence, we only need to consider the case of $C^*<2p_{\pi(n)}$. Thus no two jobs from the set $ \{J_1, J_2, \dots, J_{\pi(n)}\} $ can be scheduled on the same machine in $\sigma^*$. We have $ J_i \in J_i^* $ for any $ 1\leq i \leq \pi(n) $ by Lemma \ref{l5}(ii). Moreover, $\pi(n) \leq m$ and thus $n \leq m + \tau$ by Lemma \ref{l0}(i).
 	
If $\tau < m$, then $n <2m$ and ${\cal J}_1^*=\{J_{1}\}$ by Lemma \ref{l5}(iii).  By Lemma \ref{l0}(ii), $\pi^{-1}(1) \leq \tau+1 \leq m$. By Lemma \ref{l3}, $J_1$ is the first and  unique job scheduled on $M_{\pi^{-1}(1)}$ in $\sigma^P$. Hence, by Lemma \ref{l4}, $\frac{C^P}{C^*} \leq r_{\tau, m-1}\leq r_{\tau, m}$. We assume that $\tau \geq m$ in the following.

If $s_n \geq p_{\pi(n)}$, then by (\ref{e2}),
$$\frac{C^P}{C^*} \leq \frac{s_n + p_{\pi(n)}}{s_n + \frac{1}{m} p_{\pi(n)}} \leq \frac{p_{\pi(n)}+ p_{\pi(n)}}{p_{\pi(n)} + \frac{1}{m} p_{\pi(n)}} =\frac{2m}{m + 1} \leq \frac{2\tau}{\tau + 1} = 2 - \frac{2}{\tau + 1}\leq 2 - \frac{1}{\left\lfloor \frac{\tau}{2} \right\rfloor + 1}.$$	
If $s_n <p_{\pi(n)}$ and $ C^* \geq \frac{1}{\tau} s_n + p_{\pi(n)} $, then 
$$\frac{C^P}{C^*} \leq \frac{s_n + p_{\pi(n)}}{\frac{1}{\tau} s_n + p_{\pi(n)} } \leq \frac{p_{\pi(n)} + p_{\pi(n)}}{\frac{1}{\tau} p_{\pi(n)} + p_{\pi(n)} }
= \frac{2\tau}{\tau + 1} = 2 - \frac{2}{\tau + 1}\leq 2 - \frac{1}{\left\lfloor \frac{\tau}{2} \right\rfloor + 1}.$$
For the remaining case of $s_n <p_{\pi(n)}$ and $ C^* < \frac{1}{\tau} s_n + p_{\pi(n)} $, a more detailed comparison and analysis of the jobs on each machine in $\sigma^P$ and $\sigma^*$ is required.

We partition the first $\pi(n)$ machines into two subsets, ${\cal M}_X$ and ${\cal M}_Y$,  based on the number of jobs they process in $\sigma^*$. ${\cal M}_X$ contains machines processing exactly one job, and ${\cal M}_Y$ contains machines processing at least two jobs. We partition the last $m-\pi(n)$ machines into three subsets, ${\cal M}_U$, ${\cal M}_V$, and ${\cal M}_W$. The subsets ${\cal M}_U$ and ${\cal M}_V$ consist of machines processing exactly one job in $\sigma^*$, distinguished by whether the processing time of the job is greater than 
$s_n$ or is at most $s_n$, respectively. The subset ${\cal M}_W$ consists of machines processing at least two jobs in $\sigma^*$ (Ref. Figure \ref{fig2}). Denote by $\mu_Z=|{\cal M}_Z|$ for $Z\in\{X,Y, U, V, W\}$. Then $\pi(n)=\mu_X+\mu_Y$
and 
\begin{equation}\label{e3}
	\pi(n)+\tau\geq n \geq \mu_X+2\mu_Y+\mu_U+\mu_V+2\mu_W\geq \pi(n)+\mu_Y+2\mu_W.
\end{equation}
By Lemma \ref{l5}(iii), ${\cal M}_X = \{M_1, \ldots, M_{\mu_X}\}$, and ${\cal M}_U = {\cal M}_V= \emptyset$ if ${\cal M}_Y \neq \emptyset$. 

\begin{figure}
	\includegraphics[width=180mm]{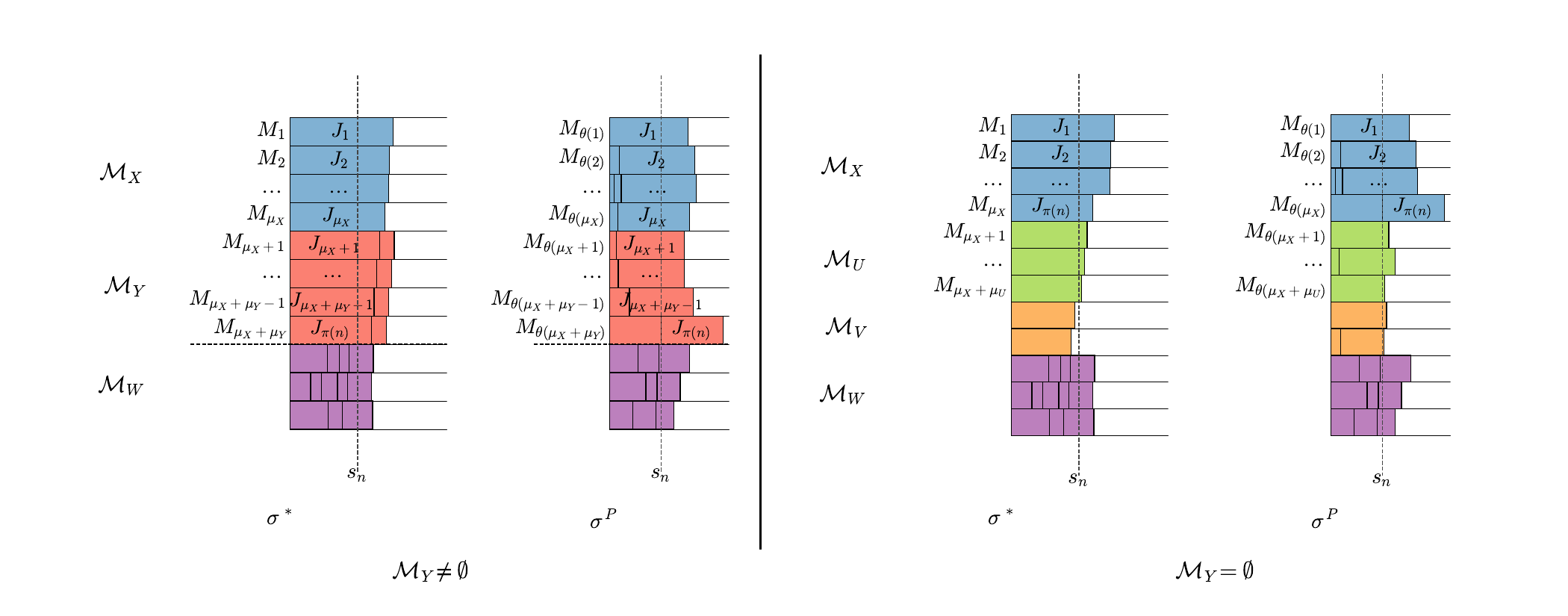}
	\centering
	\caption{Machine classification and mapping between the first $\pi(n)$ machines in $\sigma^*$ and $\sigma^P$. }\label{fig2}
\end{figure}

For any $M_i\in {\cal M}_X \cup {\cal M}_Y$, $J_i\in \mathcal{J}_{i}^{*}$ and $p_i\geq p_{\pi(n)}>s_n$. For any $M_i\in {\cal M}_U$, $J_i\in \mathcal{J}_{i}^{*}$ and $p_i>s_n$. 
Define a bijection $\theta$ on ${\cal M}$ such that for any $M_i \in {\cal M}_X \cup {\cal M}_Y \cup {\cal M}_U$, $J_i\in \mathcal{J}_{\theta(i)}^{P}$. 
By Lemma \ref{l7}, the bijection is well-defined. Regarding the loads of the machines in different sets in $\sigma^*$, and the loads of their corresponding machines in $\sigma^P$, we have the following estimates.
\begin{proposition}\label{t12}
(i) If $M_{\pi(n)} \in \mathcal{M}_X$, then $L_{\pi(n)}^* = p_{\pi(n)}$ and $L_{\theta(\pi(n))}^P = s_n + p_{\pi(n)}$.

(ii) If $M_{\pi(n)} \in \mathcal{M}_Y$, then $L_{\pi(n)}^* < \frac{1}{\tau} s_n + p_{\pi(n)}$ and $L_{\theta(\pi(n))}^P = s_n + p_{\pi(n)}$.

(iii) For any $M_i \in \mathcal{M}_X \setminus \{M_{\pi(n)}\}$, $L_i^* = p_i \leq L_{\theta(i)}^P$.

(iv) For any $M_i \in \mathcal{M}_Y \setminus \{M_{\pi(n)}\}$, $L_i^* < \frac{1}{\tau} s_n + p_{\pi(n)}$ and $L_{\theta(i)}^P \geq p_i \geq p_{\pi(n)}$.

(v) For any $M_i \in \mathcal{M}_U$, $L_i^* = p_i \leq L_{\theta(i)}^P$.

(vi) For any $M_i \in \mathcal{M}_V$, $L_i^* \leq s_n \leq L_{\theta(i)}^P$.

(vii) For any $M_i \in \mathcal{M}_W$, $L_{\theta(i)}^P \geq s_n$. 	
\end{proposition}

By Proposition \ref{t12}, regardless of whether $M_{\pi(n)} \in \mathcal{M}_X$ or $M_{\pi(n)} \in \mathcal{M}_Y$, we have
\begin{align*}
	\sum_{M_i \in {\cal M}_W} \left( L_i^*-L_{\theta(i)}^P\right)  &= \sum_{M_i \notin {\cal M}_W} \left(L_{\theta(i)}^P- L_i^*\right) \\
	&= \sum_{M_i \in {\cal M}_X \setminus \{M_{\pi(n)}\}} (L_{\theta(i)}^P - L_i^*) + \sum_{M_i \in {\cal M}_Y \setminus \{M_{\pi(n)}\}} (L_{\theta(i)}^P - L_i^*)  \\
	&\ \  + (L_{\theta(\pi(n))}^P - L_{\pi(n)}^*)+ \sum_{M_i \in {\cal M}_U} (L_{\theta(i)}^P - L_i^*) + \sum_{M_i \in {\cal M}_V} (L_{\theta(i)}^P - L_i^*)  \\
	&\geq s_n - \frac{1}{\tau} \mu_Y s_n,
\end{align*}	
and thus
\begin{align}\label{e4}
	C^* &\geq \frac{1}{\mu_W} \sum_{M_i \in {\cal M}_W} L_i^* \geq \frac{1}{\mu_W}\left( \sum_{M_i \in {\cal M}_W} \left( L_i^*-L_{\theta(i)}^P\right)+\sum_{M_i \in {\cal M}_W}L_{\theta(i)}^P\right) \nonumber \\
	&\geq \frac{s_n - \frac{1}{\tau} \mu_Y s_n + \mu_W s_n}{\mu_W} = \frac{\mu_W - \frac{1}{\tau} \mu_Y + 1}{\mu_W} s_n.
\end{align}

If $\tau$ is an even number, then by (\ref{e4}) and (\ref{e3}),
$$
C^* \geq \frac{\mu_W - \frac{1}{\tau} \mu_Y + 1}{\mu_W} s_n \geq  \frac{\mu_W - \frac{1}{\tau}\left(\tau-2\mu_W  \right)  + 1}{\mu_W} s_n=\left(\frac{2}{\tau}+1 \right)s_n
$$
and 
$$C^P = s_n + p_{\pi(n)} \leq \frac{1}{\frac{2}{\tau}+1} C^* + C^*= \frac{2\tau + 2}{\tau + 2}C^* = \left(2 - \frac{1}{\frac{\tau}{2} + 1}\right) C^*=\left(2 - \frac{1}{\left\lfloor \frac{\tau}{2} \right\rfloor + 1}\right) C^*. $$
If $\tau$ is an odd number and $\tau \geq \mu_Y+2\mu_W+ 1\geq 2\mu_W + 1$, then by (\ref{e4}),
$$
\begin{aligned}
C^* &\geq  \frac{\mu_W - \frac{1}{\tau} \mu_Y + 1}{\mu_W} s_n \geq  \frac{\mu_W - \frac{1}{\tau}\left(\tau-2\mu_W  -1\right)  + 1}{\mu_W} s_n \\
&=\left(1+\frac{2}{\tau}+\frac{1}{\mu_W\tau}\right)s_n\geq \left(1 + \frac{2}{\tau} + \frac{1}{\frac{\tau - 1}{2}\tau }\right) s_n = \frac{\tau + 1}{\tau - 1} s_n
\end{aligned}
$$
and 
$$C^P = s_n + p_{\pi(n)} \leq \frac{\tau-1}{\tau+1} C^* + C^*= \frac{2\tau}{\tau + 1}C^* = \left(2 - \frac{1}{\frac{\tau-1}{2} + 1}\right) C^*=\left(2 - \frac{1}{\left\lfloor \frac{\tau}{2} \right\rfloor + 1}\right) C^*. $$
If $\tau$ is an odd number and $\tau =\mu_Y+2\mu_W$, then $\mu_Y\geq 1$. Thus $M_{\pi(n)}\in {\cal M}_Y$ and $\mu_U=\mu_V=0$. Moreover,  $\pi(n)=\mu_X+\mu_Y=n-\tau$ and $|{\cal J}_i^*| = 2$ for any $M_i \in {\cal M}_Y \cup {\cal M}_W$ by (\ref{e3}). Since $\mu_1^*=\mu_X$ and $\mu_2^*=\mu_Y+\mu_W$, $n=2m-\mu_X=2m-\pi(n)+\mu_Y$ by \eqref{t11}, and thus by Lemma \ref{l5}(iv),
\begin{equation}\label{t74}
	C^*\geq L_{\pi(n)}^*=p_{\pi(n)}+p_{2m+1-{\pi(n)}}=p_{\pi(n)}+p_{n - \mu_Y + 1}.
\end{equation}
Therefore, by Lemma \ref{l5}(iv) and Proposition \ref{t12}(ii)(iv), for any $\mu_X+1\leq i\leq \mu_X+\mu_Y-1$,
\begin{equation}\label{t14}
L_{\theta(i)}^P - L_i^*\geq p_i-(p_i+p_{2m+1-i})=-p_{2m+1-i}\geq -p_{2m+1-(\mu_X+\mu_Y)}=- p_{n - \mu_Y + 1}	
\end{equation}
and
\begin{equation}\label{t15}
	L_{\theta(\pi(n))}^P - L_{\pi(n)}^*\geq s_n+p_{\pi(n)}-(p_{\pi(n)}+p_{n - \mu_Y + 1})=s_n- p_{n - \mu_Y + 1}.	
\end{equation}
Moreover,
\begin{align}\label{e5}
	C^* &=(\tau+1)C^*-\tau C^*\geq (\tau+1)(p_{\pi(n)} + p_{n - \mu_Y + 1})-\tau\left( \frac{1}{\tau}s_n+ p_{\pi(n)}\right)  \nonumber \\
	&=p_{\pi(n)}-s_n+(\tau+1)p_{n - \mu_Y + 1}>  (\tau+1) p_{n - \mu_Y + 1}.
\end{align}
By \eqref{t14}, \eqref{t15} and Proposition \ref{t12}(iii)(vii),
\begin{align*}\label{e7}
	\mu_W C^* &\geq \sum_{M_i \in {\cal M}_W} L_i^* = \sum_{M_i \notin {\cal M}_W} \left(L_{\theta(i)}^P- L_i^*\right) +\sum_{M_i \in {\cal M}_W} L_{\theta(i)}^P \nonumber\\
	&= \sum_{i=1}^{\mu_X} (L_{\theta(i)}^P - L_i^*) + \sum_{i=\mu_X+1}^{\mu_X+\mu_Y-1}  (L_{\theta(i)}^P - L_i^*) + (L_{\theta(\pi(n))}^P - L_{\pi(n)}^*) \nonumber + \sum_{i=\mu_X+\mu_Y+1}^{\mu_X+\mu_Y+\mu_W} L_{\theta(i)}^P \nonumber\\
	& 
	\geq s_n-\mu_Y p_{n-\mu_Y +1}+\mu_W s_n= \left(\mu_W  + 1\right) s_n-\mu_Y p_{n-\mu_Y +1}.
\end{align*}	
Combining it with \eqref{t74}, we have
\begin{align*}
 (2\mu_W+1) C^* &= \mu_W C^*+(\mu_W+1) C^* \\
 & \geq \left(\mu_W  + 1\right) s_n-\mu_Y p_{n-\mu_Y +1}+(\mu_W+1)(p_{\pi(n)} + p_{n - \mu_Y + 1}) \\
 &=	(\mu_W+1)(s_n+p_{\pi(n)}) + (\mu_W-\mu_Y+1)p_{n - \mu_Y + 1}
\end{align*}
and thus
$$C^P = s_n + p_{\pi(n)} \leq \frac{2\mu_W+1}{\mu_W+1} C^* - \frac{\mu_W-\mu_Y+1}{\mu_W+1}p_{n - \mu_Y + 1}. $$
If $\mu_W-\mu_Y+1\geq 0$, then 
$$C^P = s_n + p_{\pi(n)} \leq \frac{2\mu_W+1}{\mu_W+1} C^*=\left(2 - \frac{1}{\mu_W + 1}\right) C^*\leq\left(2 - \frac{1}{\left\lfloor \frac{\tau}{2} \right\rfloor + 1}\right) C^*. $$
Otherwise, by (\ref{e5}), 
\begin{align*}
C^P &= s_n + p_{\pi(n)} \leq \frac{2\mu_W+1}{\mu_W+1} C^*-\frac{\mu_W-\mu_Y+1}{(\mu_W+1)(\tau+1)}C^*=\left(2 - \frac{\mu_W-\mu_Y+\tau+2}{(\mu_W+1)(\tau+1)}\right) C^* \\
&=\left(2 - \frac{3\mu_W+2}{(\mu_W+1)(\tau+1)}\right) C^*\leq \left(2 - \frac{2}{\tau + 1}\right) C^*
=\left(2 - \frac{1}{\left\lfloor \frac{\tau}{2} \right\rfloor + 1}\right) C^*. 
\end{align*}

\end{proof}		

\begin{theorem}\label{thm0}
The competitive ratio of POLPT is 
\begin{equation*}\label{e0}
	r_{\tau,m}^P=\left\{ 
	\begin{array}{ll}
		\frac{4m-1}{3m} & \tau=0, \\
		\max\{\frac{5}{4},\frac{4m-2}{3m-1}\} & \tau= 1, \\
		2 - \frac{1}{\left\lfloor \frac{\tau}{2} \right\rfloor + 1} & 2\leq \tau \leq 2m-2, \\
		2 - \frac{1}{m} & \tau > 2m-2,  
	\end{array} 
	\right.	
\end{equation*} 
and the bound is tight for any $\tau$ and $m$.	
\end{theorem}
\begin{proof}
Note that when $\tau=0$, POLPT reduces to LPT, yielding an identical schedule. From the conclusion on the worst-case ratio of LPT in \cite{Graham69}, the result follows directly. The upper bounds for the remaining cases can be derived from  the previous discussion and Lemmas \ref{l2}(i), \ref{m2}, and \ref{m3}. Finally, we provide instances to demonstrate the tightness of the bound.
	
For $\tau=1$ and $m=2$, consider an instance $I_1$ with $4$ jobs. The processing time of the jobs are $p_1=3, p_2=p_3=2, p_4=1$. Permutation $\pi$  is given by (\ref{e9}). Obviously, ${\cal J}_1^P=\{J_1,J_3\}$ and ${\cal J}_2^P=\{J_2,J_4\}$, while ${\cal J}_1^*=\{J_1,J_4\}$ and ${\cal J}_2^*=\{J_2,J_3\}$. Thus $C^P=5$, $C^*=4$ and $\frac{C^P(I_1)}{C^*(I_1)}=\frac{5}{4}$.
	
For $\tau=1$  and $m\geq 3$, consider an instance $I_2$ with $2m+1$ jobs. The processing time of the jobs are 
$$p_1=2m-1, \quad p_{2j}=p_{2j+1}=2m-1-j, j=1,\dots, m-1, \quad p_{2m}=m, \quad p_{2m+1}=m-1.$$ 
Permutation $\pi$  is given by (\ref{e9}). It is not difficult to verify that 
$${\cal J}_1^P=\{J_1,J_{2m+1},J_{2m}\}, \quad {\cal J}_i^P=\{J_i,J_{2m+1-i}\}, i=2,\dots,m.$$ Thus $C^P=4m-2$. On the other hand, 
$${\cal J}_i^*=\{J_{i},J_{2m-1-i}\}, i=1,\dots,m-1, \quad  {\cal J}_m^*=\{J_{2m-1},J_{2m},J_{2m+1}\}.$$  
Thus $C^*=3m-1$ and $\frac{C^P(I_2)}{C^*(I_2)}= \frac{4m-2}{3m-1}$.

For $2\leq \tau\leq 2m-2$, consider an instance $I_3$ with $m+\left\lfloor \frac{\tau}{2} \right\rfloor$ jobs. The processing time of the jobs are 
$$p_1=\dots=p_{m-\left\lfloor \frac{\tau}{2} \right\rfloor}=\left\lfloor \frac{\tau}{2} \right\rfloor+1, \quad p_{m-\left\lfloor \frac{\tau}{2} \right\rfloor+1}=\dots=p_{m}=\left\lfloor \frac{\tau}{2} \right\rfloor, \quad  p_{m+1}=\dots=p_{m+\left\lfloor \frac{\tau}{2} \right\rfloor}=1.$$ 
Permutation $\pi$  is given by (\ref{e9}). It is not difficult to verify that 
\begin{align*}
	& {\cal J}_i^P=\{J_i\}, i=1,\dots,m-\left\lceil \frac{\tau}{2} \right\rceil-1, \quad {\cal J}_i^P=\{J_{i+1}\}, i=m-\left\lceil \frac{\tau}{2} \right\rceil,\dots,m-1, \\
	& {\cal J}_m^P=\{J_{m+1},\dots, J_{m+\left\lfloor \frac{\tau}{2} \right\rfloor}, J_{m-\left\lceil \frac{\tau}{2} \right\rceil}\}.	
\end{align*}
Thus $C^P=2\left\lfloor \frac{\tau}{2} \right\rfloor+1$. On the other hand, 
$${\cal J}_i^*=\{J_i\}, i=1,\dots,m-\left\lfloor \frac{\tau}{2} \right\rfloor, \quad {\cal J}_i^*=\{J_{i},J_{2m+1-i}\}, i=m-\left\lfloor \frac{\tau}{2} \right\rfloor+1,\dots,m.$$  Thus $C^*=\left\lfloor \frac{\tau}{2} \right\rfloor+1$ and $\frac{C^P(I_3)}{C^*(I_3)}= \frac{2\left\lfloor \frac{\tau}{2} \right\rfloor+1}{\left\lfloor \frac{\tau}{2} \right\rfloor+1}=2 - \frac{1}{\left\lfloor \frac{\tau}{2} \right\rfloor + 1}$.	
Note that $2 - \frac{1}{\left\lfloor \frac{\tau}{2} \right\rfloor + 1}=2-\frac{1}{m}$ when $\tau=2m-2$. Therefore, this instance also implies that the competitive ratio of the algorithm is no less than $2-\frac{1}{m}$ when $\tau >2m-2$.
\end{proof}


\section{Optimal Algorithm for Three Machines}

In this section, we present an optimal algorithm DLPT (Deferred LPT) for three machines when the prediction is completely correct. This algorithm achieves better performance than POLPT by making special assignments for the $7$ jobs with the largest processing times. The algorithm also applies to instances with $n < 7$ by setting $p_{n+1}=\ldots=p_{7}=0$.

Let $\alpha=\frac{5+\sqrt{15}}{5}\approx 1.775 $ be the unique solution in the interval $(1,\infty)$ of the equation $\frac{2x+1}{x+2}=\frac{7x-2}{6x-2}$. Let 
\begin{equation}\label{t71}
\beta=\frac{2\alpha+1}{\alpha+2}=\frac{7\alpha-2}{6\alpha-2}=\frac{13+\sqrt{15}}{14}\approx 1.205	
\end{equation}
be the competitive ratio of the DLPT, which will be shown later.

\textbf{Algorithm DLPT}

\begin{enumerate}
	\item Assign $J_1$ to $M_1$. Assign $J_7$ and $J_2$ to $M_2$, processing $J_7$ before $J_2$. Assign $J_3$ to $M_3$.
	
	\item If $p_3 \geq \alpha p_7$, assign $J_4$ to $M_3$. Otherwise, assign $J_5$ to $M_3$.
	
	\item Assign the remaining jobs one by one in increasing order of their indices to the machine where they can be started earliest.
\end{enumerate}

Note that Step 1 can be executed at time $0$.  Step 2 can be executed when $J_3$
completes, and Step 3 is executed whenever a machine becomes idle and there are remaining jobs to be processed. 

Next we present some lower bounds on the optimum for various situations. By Lemma \ref{l5}, we assume w.l.o.g. that $J_1\in \mathcal{J}_1^*$,  and $J_2\in \mathcal{J}_2^*$ if $J_2\notin \mathcal{J}_1^*$. Moreover, $|\mathcal{J}_2^*|=1$ only if $|\mathcal{J}_1^*|=1$. 

\begin{lemma}\label{tho}
(i) $C^* \geq \frac{P}{3}$ and $C^* \geq \max\{p_1, p_3+p_4, p_5+p_6+p_7\}$.

(ii) If $\mathcal{J}_1^* = \{J_1\}$, then $C^* \geq \frac{1}{2}(P - p_1) \geq p_3+p_5+p_7$ and $C^* \geq 3p_6$.

(iii)  If each machine processes at least two jobs of $J_1, \dots, J_7$ in $\sigma^*$, then $C^* \geq p_1+p_7$ and $C^* \geq p_2+p_6$.

(iv) If $\mathcal{J}_1^* = \{J_1,J_7\}$, then $C^* \geq 3p_6$ and $C^* \geq \frac{2}{3}(p_3+p_4+2p_6)$.

(v) If each machine processes at most three jobs of $J_1, \dots, J_8$ in $\sigma^*$, then $C^* \geq p_1+ p_8$ and $C^* \geq p_3 + p_7 + p_8$. If in addition $J_2 \notin \mathcal{J}_1^*$, then $C^* \geq p_2 + p_7 + p_8$.
\end{lemma}

\begin{proof}
(i) Obviously, $C^* \geq \frac{P}{3}$ and $C^* \geq p_1$. Since there must be a machine which processes at least two of $J_1,\dots, J_4$, $C^* \geq p_3 + p_4$. Similarly, since there must be a machine which processes at least three of $J_1, \dots, J_7$, $C^* \geq p_5+p_6+p_7$.
	
(ii) Clearly, 
\begin{align*}
C^*&=\max\limits_{i=1,2,3}P(\mathcal{J}_i^*) \geq \frac{1}{2}\left( P(\mathcal{J}_2^*) + P(\mathcal{J}_3^*) \right)=\frac{1}{2}\left( P - p_1 \right) \\
& \geq \frac{1}{2}\left( p_2 + p_3 + p_4 + p_5 + p_6 + p_7 \right) \geq p_3 + p_5 + p_7.	
\end{align*}
If there exists $i>1$ such that $M_i$ processes at least four jobs of $J_2,\dots, J_7$ in $\sigma^*$, then at least three jobs of $\mathcal{J}_i^*$ have a processing time of at least $p_6$. Otherwise, each of $M_2$ and $M_3$ processes exactly three jobs of $J_2,\dots,J_7$ in $\sigma^*$, and exactly one of $\mathcal{J}_2^*$ and $\mathcal{J}_3^*$ does not contain $J_7$. In both cases, $C^* \geq 3p_6$.
	
(iii) If $\mathcal{J}_1^* = \{J_1, J_2\}$, then the result clearly follows. Otherwise, $J_1$ and $J_2$ are each processed on a machine with at least one other job of $J_3,\dots,J_7$. Hence, $C^* \geq p_1+p_7$ and $C^* \geq p_2+p_6$.

(iv) Since there must be a machine, say $M_l$, among $M_2$ and $M_3$ which processes at least three jobs of $J_2,\dots, J_6$, $C^* \geq 3p_6$. Moreover,
\begin{align*}
	C^*&=\max\limits_{i=1,2,3}P(\mathcal{J}_i^*) \geq \frac{1}{3}\left( P(\mathcal{J}_2^*) + P(\mathcal{J}_3^*) \right)+ \frac{1}{3}P(\mathcal{J}_l^*)=\frac{1}{3}\left( P - p_1-p_7 \right)+\frac{1}{3}P(\mathcal{J}_l^*) \\
	& \geq \frac{1}{3}\left( p_2 + p_3 + p_4 + p_5 + p_6\right)+\frac{1}{3}(p_4 + p_5 + p_6)\geq \frac{2}{3}(p_3+p_4+2p_6).	
\end{align*}

(v) There are two machines which process three jobs of $J_1, \dots, J_8$ in $\sigma^*$. If $|\mathcal{J}_1^*| = 3$, then the result follows immediately. Otherwise, $|\mathcal{J}_1^*| = 2$ and thus $C^* \geq p_1+ p_8$. Moreover, in $\sigma^*$, the jobs among $J_2$ and $J_3$ that are not processed on $M_1$ is processed on the same machine as two other jobs. Thus, $C^* \geq p_3 + p_7 + p_8$, and if $J_2 \notin \mathcal{J}_1^*$, $C^* \geq p_2 + p_7 + p_8$.
\end{proof}

\begin{figure}
	\includegraphics[width=160mm]{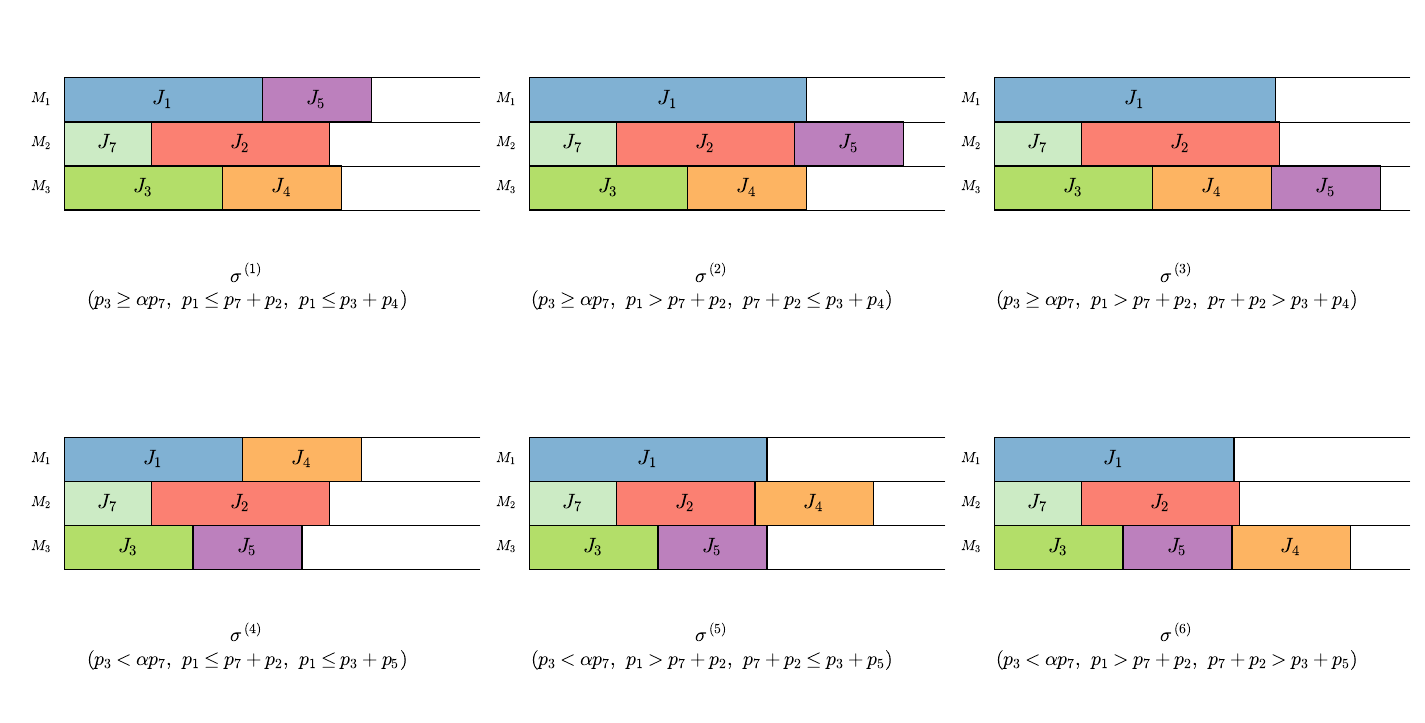}
	\centering
	\caption{Partial schedules of $\sigma^L$. }\label{fig3}
\end{figure}

The rest of this section proves that the competitive ratio of Algorithm DLPT is at most $\beta$. 
Let the schedule generated by the DLPT be denoted by $\sigma^L$. Based on the description of the algorithm, we can enumerate all possible partial schedules of the six jobs with the earliest start times, namely $J_1, J_2, J_3, J_4, J_5$ and $J_7$, which are denoted as $\sigma^{(1)}, \sigma^{(2)}, \dots, \sigma^{(6)}$. (Ref. Figure \ref{fig3}) Each partial schedule occurs under specific conditions. For example, if the partial schedule of $\sigma^L$ is one of $\sigma^{(k)}$, where $k=4,5,6$, then
\begin{equation}\label{t72}
	p_6\leq p_5\leq p_4\leq p_3<\alpha p_7.
\end{equation}
Specifically, if the partial schedule of $\sigma^L$ is $\sigma^{(4)}$, we further have
\begin{equation}\label{t73}
p_1 \leq \min\{p_2 + p_7, p_3 + p_5\}.
\end{equation}

\begin{lemma}\label{thr2}
	If $n\geq 8$, then $\frac{C^L}{C^*} \leq \beta$.
\end{lemma}

\begin{proof}
	If $J_n$ dost not determine the makespan of $\sigma^L$, then $C^L(I_{n-1}) = C^L(I)$ as $n\geq 8$. Since $C^*(I_{n-1}) \leq C^*(I)$, we have $\frac{C^L(I)}{C^*(I)} \leq \frac{C^L(I_{n-1})}{C^*(I_{n-1})}$. The subsequent Lemma \ref{thr3} proves that $\frac{C^L}{C^*} \leq \beta$  when $n\leq 7$. Hence, to prove the competitive ratio of the DLPT, it is sufficient to consider the instances that $J_{n}$ determines the makespan. 
	
	Since $J_n$ is scheduled to the machine where it can be started the earliest, all the machines would not be idle before the start time of $J_{n}$. Similarly as \eqref{e1}, 
	\begin{equation}\label{e30}
		C^L \leq \frac{1}{3}(P - p_n) + p_n = \frac{1}{3}P + \frac{2}{3}p_n \leq C^* + \frac{2}{3}p_n.
	\end{equation}
	If there exists a machine which processes at least four jobs in $\sigma^*$, then $C^* \geq 4p_n$. Thus, by (\ref{e30}), $\frac{C^L}{C^*} \leq 1 + \frac{1}{6} = \frac{7}{6} \leq \beta$.
	Therefore, we assume in the following that each machine processes at most three jobs in $\sigma^*$. Then $n \leq 9$. 
	
	If $n=9$, then obviously $C^* \geq p_1 + p_8 + p_9$. If $p_1 \geq \frac{4}{3}p_9$, then  $C^* \geq p_1 + p_8 + p_9 \geq \frac{10}{3}p_9$. By \eqref{e30},
	\[
	\frac{C^L}{C^*} \leq 1 + \frac{\frac{2}{3}p_9}{\frac{10}{3}p_9} = 1 + \frac{1}{5} = \frac{6}{5} \leq \beta.
	\]
	Otherwise, $p_3 \leq p_1 < \frac{4}{3}p_9 < \alpha p_7< \min\{p_2 + p_7, p_3 + p_5\}$. Hence, the partial schedule of $\sigma^L$ is $\sigma^{(4)}$ by \eqref{t72} and \eqref{t73}. Since $p_1 + p_4 \geq \max\{p_2 + p_7, p_3 + p_5\}$, neither $J_6$ nor $J_8$ is processed on $M_1$. Therefore, $C^L \leq p_1 + p_4 + p_9$, and by $p_1 < \frac{4}{3}p_9$,
	\[
	\frac{C^L}{C^*} \leq \frac{p_1 + p_4 + p_9}{p_1 + p_8 + p_9} = 1 + \frac{p_4 - p_8}{p_1 + p_8 + p_9} \leq 1 + \frac{p_1 - p_9}{3p_9} \leq 1 + \frac{1}{9} = \frac{10}{9} \leq \beta.
	\]
	
	We turn to the case of $n=8$. If $p_3 \geq \frac{4}{3}p_8$, then by Lemma \ref{tho}(v), $C^* \geq p_3 + p_7 + p_8 \geq \frac{10}{3}p_8$. By \eqref{e30},
	\[
	\frac{C^L}{C^*} \leq 1 + \frac{\frac{2}{3}p_8}{\frac{10}{3}p_8} \leq 1 + \frac{1}{5} = \frac{6}{5} \leq \beta.
	\]
	If $p_3 < \frac{4}{3}p_8 \leq \alpha p_7$, then the partial schedule of $\sigma^L$ is one of $\sigma^{(k)}$, $k=4, 5, 6$. Note that there is at least one machine which does not process $J_4$ or $J_6$. We consider the cases depending on which machine it is.
	
	If $M_1$ does not process $J_4$ or $J_6$, then $C^L \leq p_1 + p_8 \leq C^*$. If $M_2$ does not process $J_4$ or $J_6$, then $C^L \leq p_7 + p_2 + p_8$. If $J_1$ and $J_2$ are processed on distinct machines in $\sigma^*$, then $C^* \geq p_2 + p_7 + p_8 \geq C^L$ by  Lemma \ref{tho}(v). Otherwise, $C^* \geq p_1 + p_2$ and $C^* \geq p_3 + p_7 + p_8$ by  Lemma \ref{tho}(v). Therefore,
	\[
	\frac{C^L}{C^*} \leq \frac{p_2 + p_7 + p_8}{\max\{p_1 + p_2, p_3 + p_7 + p_8\}} \leq \frac{p_2 + \frac{2}{3}(2p_7 + p_8)}{\max\{2p_2, 2p_7 + p_8\}} \leq \frac{1}{2} + \frac{2}{3} = \frac{7}{6} \leq \beta.
	\]
	If $M_3$ does not process $J_4$ or $J_6$, then $C^L \leq p_3 + p_5 + p_8$. Therefore, by Lemma \ref{tho}(v) and $p_3 < \frac{4}{3}p_8$,
	\[
	\frac{C^L}{C^*} \leq \frac{p_3 + p_5 + p_8}{p_3 + p_7 + p_8} = 1 + \frac{p_5 - p_7}{p_3 + p_7 + p_8} \leq 1 + \frac{p_3 - p_8}{3p_8} \leq 1 + \frac{1}{9} = \frac{10}{9} \leq \beta.
	\]
\end{proof}

We now consider the case $n\leq 7$. Note that in this case, $J_n$ is not necessarily the job that determines the makespan of $\sigma^L$, nor is the assignment of jobs $J_1, \ldots, J_{n-1}$ in $\sigma^L(I_{n-1})$ necessarily identical to that in $\sigma^L(I_{n})$. Therefore, a detailed discussion of each possible schedule seems unavoidable.

\begin{lemma}\label{thr3}
	If $n\leq 7$, then $\frac{C^L}{C^*} \leq \beta$.
\end{lemma}
\begin{proof}
We analyze the cases based on which job determines the makespan of schedule $\sigma^L$. Note that job $J_3$ or $J_7$ cannot be the job determining the makespan. If $J_1$ determines the makespan of $\sigma^L$, then obviously $C^L = p_1 \leq C^*$. If $J_2$ determines the makespan of $\sigma^L$, then $C^L = p_7+p_2$. If $J_2$ is not scheduled alone on a machine in $C^*$, then $C^* \geq p_2+p_7 \geq C^L$. Otherwise, $|\mathcal{J}_1^*|=|\mathcal{J}_2^*|=1$ and thus $C^* \geq p_3+p_4+p_5+p_6+p_7 \geq 5p_7$. Thus 
\[
\frac{C^L}{C^*} \leq \frac{p_2+p_7}{\max\{p_2, 5p_7\}} \leq 1 + \frac{1}{5} = \frac{6}{5}\leq \beta.
\]

If $J_6$ determines the makespan of $\sigma^L$, by the rule of the algorithm,
\begin{equation}\label{e33}
C^L \leq \min\{L_2^{(i)}, L_3^{(i)}\} + p_6 \leq \frac{1}{2}(L_2^{(i)} + L_3^{(i)}) + p_6 \leq \frac{1}{2}(p_2+p_3+p_4+p_5+p_7) + p_6 = \frac{1}{2}(P-p_1) + \frac{1}{2}p_6,
\end{equation}
where $L^{(k)}_{i}$ is the load of machine $M_i$ in partial schedule $\sigma^{(k)}$ for $i=1,2,3$ and $k=1,\dots, 6$.

If $\mathcal{J}_1^* = \{J_1\}$, by \eqref{e33} and Lemma \ref{tho}(ii),
\[
\frac{C^L}{C^*} \leq \frac{\frac{1}{2}(P-p_1) + \frac{1}{2}p_6}{\max\{\frac{1}{2}(P-p_1), 3p_6\}} \leq 1 + \frac{1}{6} = \frac{7}{6} \leq \beta.
\]
If $|\mathcal{J}_i^*|\geq 2$ for any $i$, we consider different cases based on which of $\sigma^{(k)}$, $k=1,\dots, 6$, the partial schedule $\sigma^L$ is.

If the partial schedule of $\sigma^L$ is $\sigma^{(1)}$, then
\begin{equation}\label{e35}
C^L=\min\{p_1+p_5, p_2+p_7, p_3+p_4\} + p_6 \leq \min\{p_2+p_7, p_3+p_4\} + p_6.
\end{equation}
If one of $J_1$ and $J_2$ is scheduled on the same machine as two other jobs in $\sigma^*$, then $C^* = p_2+p_6+p_7 \geq C^L$. If one of $J_1$ or $J_2$ is processed on the same machine as at least one of $J_5, J_6$ or $J_7$ in $\sigma^*$,  then 
$C^* \geq \frac{1}{2}(P-p_1-p_5)$. 
Combining the above  inequality with Lemma \ref{tho}(i)(iii) and \eqref{e35}, we have
\begin{align*}
\frac{C^L}{C^*} &\leq \frac{\frac{3}{5}(p_2+p_7)+\frac{2}{5}(p_3+p_4) + p_6}{\max\{\frac{1}{2}(P-p_1-p_5),p_2+p_6,p_5+p_6+p_7\}} \\
&= \frac{\frac{2}{5}(p_2+p_3+p_4+p_6+p_7)+\frac{1}{5}(p_2+p_6) +\frac{1}{5}(2p_6+p_7)}{\max\{\frac{1}{2}(P-p_1-p_5),p_2+p_6,p_5+p_6+p_7\}} \\
& \leq \frac{\frac{2}{5}(P-p_1-p_5)+\frac{1}{5}(p_2+p_6) +\frac{1}{5}(2p_6+p_7)}{\max\{\frac{1}{2}(P-p_1-p_5),p_2+p_6,2p_6+p_7\}}\leq \frac{6}{5} \leq \beta.	
\end{align*}	
Hence, we are left with only the case where in $\sigma^*$, $J_1, J_2, J_3$ and $J_4$ are processed on two machines, with exactly two jobs on each machine. Thus $C^* \geq p_2+p_3$. By $p_3 \geq \alpha p_7$, 
\begin{align*}
	\min\{p_2+p_7, p_3+p_4\} &\leq \frac{4\alpha-1}{6\alpha-2}(p_2+p_7) + \frac{2\alpha-1}{6\alpha-2}(p_3+p_4) \\
	&\leq \frac{4\alpha-1}{6\alpha-2}p_2 + \frac{4\alpha-2}{6\alpha-2}p_3 + \frac{\alpha}{6\alpha-2}p_7 + \frac{1}{2}p_7\leq \frac{4\alpha-1}{6\alpha-2}(p_2+p_3) + \frac{1}{2}p_7.
\end{align*}
Therefore, by Lemma \ref{tho}(i) and \eqref{t71}, we have
\begin{align*}
	\frac{C^L}{C^*} &\leq \frac{\min\{p_2+p_7, p_3+p_4\} + p_6 }{\max\{p_2+p_3,p_5+p_6+p_7\}}\leq \frac{\frac{4\alpha-1}{6\alpha-2}(p_2+p_3) + \frac{1}{2}p_7+p_6}{\max\{p_2+p_3,2p_6+p_7\}} \leq \frac{4\alpha-1}{6\alpha-2} + \frac{1}{2} = \frac{7\alpha-2}{6\alpha-2} = \beta.
\end{align*}

If the partial schedule of $\sigma^L$ is $\sigma^{(2)}$, then
\begin{equation}\label{e34}
	C^L = \min\{p_1, p_7+p_2+p_5, p_3+p_4\} + p_6.
\end{equation}
If $J_1$ is processed on the same machine as at least one job other than $J_7$ in $\sigma^*$, then $C^* \geq p_1+p_6 \geq C^L$. Otherwise, $\mathcal{J}_1^* = \{J_1, J_7\}$. By \eqref{e34} and Lemma \ref{tho}(i)(iv), we have
\begin{align*}
	\frac{C^L}{C^*} &\leq \frac{\frac{1}{5}p_1+\frac{1}{5}(p_7+p_2+p_5)+\frac{3}{5}(p_3+p_4) + p_6}{\max\{\frac{1}{3}P, \frac{2}{3}(p_3+p_4+2p_6)\}} =\frac{\frac{1}{5}P+\frac{2}{5}(p_3+p_4+2p_6)}{\max\{\frac{1}{3}P, \frac{2}{3}(p_3+p_4+2p_6)\}} \leq\frac{3}{5}+\frac{3}{5}= \frac{6}{5}\leq  \beta.
\end{align*}

If the partial schedule of $\sigma^L$ is $\sigma^{(3)}$, then
\begin{equation}\label{e36}
C^L = \min\{p_1, p_2+p_7, p_3+p_4+p_5\} + p_6\leq \min\{p_1, p_2+p_7\} + p_6.
\end{equation}
If $J_1$ is processed on the same machine as at least one job other than $J_7$ in $\sigma^*$, then $C^*\geq p_1+p_6\geq C^L$. Otherwise, $\mathcal{J}_1^* = \{J_1, J_7\}$. By \eqref{e36} and Lemma \ref{tho}(iii)(iv), we have
\[
\frac{C^L}{C^*} \leq \frac{\frac{1}{2}p_1+\frac{1}{2}(p_2+p_7) + p_6}{\max\{p_1+p_7, p_2+p_6,3p_6\}}=\frac{\frac{1}{2}(p_1+p_7)+\frac{1}{2}(p_2+p_6) + \frac{1}{2}p_6}{\max\{p_1+p_7, p_2+p_6,3p_6\}}\leq \frac{1}{2}+\frac{1}{2}+\frac{1}{6}= \frac{7}{6} \leq \beta.
\]

If the partial schedule of $\sigma^L$ is $\sigma^{(4)}$, then
\begin{equation}\label{e37}
C^L =\min\{p_1+p_4, p_7+p_2, p_3+p_5\} + p_6\leq \min\{p_7+p_2, p_3+p_5\} + p_6.
\end{equation}
If $J_1, J_2, J_3$ are processed on distinct machines in $\sigma^*$, then $C^* \geq p_3+p_6+p_7$. Therefore,  by $p_3+p_6\geq p_5+p_7$ and \eqref{t72}, \eqref{t71}, we have
\[
\frac{C^L}{C^*} \leq \frac{p_3+p_6+p_5}{p_3+p_6+p_7} \leq \frac{p_5+p_7+p_5}{p_5+p_7+p_7} = \frac{2p_5+p_7}{p_5+2p_7} \leq \frac{2\alpha+1}{\alpha+2} = \beta.
\]
If at least two of $J_1, J_2, J_3$ are processed on the same machine in $\sigma^*$, then $C^* \geq p_2+p_3$. Combining the above inequality with Lemma \ref{tho}(i), and \eqref{e37}, \eqref{t72}, we have
\begin{align*}
	\frac{C^L}{C^*} &
	\leq \frac{\frac{1}{2}(p_2+p_7)+\frac{1}{2}(p_3+p_5) + p_6}{\max\{p_2+p_3, p_5+p_6+p_7\}} =\frac{\frac{1}{2}(p_2+p_3) + \frac{1}{2}(p_5+p_6+p_7) + \frac{1}{2}p_6}{\max\{p_2+p_3, p_5+p_6+p_7\}} \\
	&\leq 1 + \frac{p_6}{2(p_5+p_6+p_7)} \leq  1 + \frac{p_6}{4p_6+2p_7} \leq 1 + \frac{\alpha p_7}{4 \alpha p_7+2p_7} = \frac{5\alpha+2}{4\alpha+2} \leq  \frac{6}{5} \leq \beta.
\end{align*}

If the partial schedule of $\sigma^L$ is $\sigma^{(5)}$ or $\sigma^{(6)}$, then $C^L \leq p_1+p_6$. By Lemma \ref{tho}(i)(iii) and \eqref{t72}, \eqref{t71}, we have
\begin{align*}
	\frac{C^L}{C^*} &\leq \frac{p_1+p_6}{\max\{p_1+p_7, p_5+p_6+p_7\}} \leq \frac{p_1+p_7+p_6-p_7}{\max\{p_1+p_7, p_6+2p_7\}} \leq 1 + \frac{p_6-p_7}{p_6+2p_7}  \\
	&\leq 1 + \frac{\alpha p_7-p_7}{\alpha p_7+2p_7} = \frac{2\alpha+1}{\alpha+2} = \beta.
\end{align*}

It remains to consider the case where either $J_4$ or $J_5$
determines the makespan of $\sigma^L$. We consider two cases depending on whether  $p_3 \geq \alpha p_7$.

If $p_3 \geq \alpha p_7$, the partial schedule of $\sigma^L$ is one of $\sigma^{(k)}$, where $k=1,2,3$. If $J_4$ determines the makespan of $\sigma^L$, then $C^L \leq  p_3+p_4 \leq C^*$ by Lemma \ref{tho}(i). If $J_5$ determines the makespan of $\sigma^L$, since $J_5$ is assigned to the machine where it can be started earliest, then
\begin{equation}\label{e31}
C^L = \min\{p_1, p_2+p_7, p_3+p_4\} + p_5.
\end{equation}

Note that if $p_2+p_7\leq p_3+p_4$, then 
$$\frac{p_2+p_7+ p_5}{(p_2+p_7+p_3+p_4)+ (p_5+p_6)} \leq \frac{p_3+p_4+ p_5}{2(p_3+p_4)+ (p_5+p_6)}.$$
Otherwise, 
$$\frac{p_3+p_4+ p_5}{(p_2+p_7+p_3+p_4)+ (p_5+p_6)} \leq \frac{p_3+p_4+ p_5}{2(p_3+p_4)+ (p_5+p_6)}.$$
Therefore, if $\mathcal{J}_1^* = \{J_1\}$, then by Lemma \ref{tho}(ii),
\begin{align*}
\frac{C^L}{C^*} &\leq \frac{\min\{p_2+p_7, p_3+p_4\} + p_5}{\frac{1}{2}(P-p_1)}=\frac{2(\min\{p_2+p_7, p_3+p_4\} + p_5)}{(p_2+p_7+p_3+p_4)+ (p_5+p_6)} \leq \frac{2(p_3+p_4+ p_5)}{2(p_3+p_4)+ (p_5+p_6)}\\
	& = 1 + \frac{p_5-p_6}{2p_3+2p_4+p_5+p_6} \leq 1 + \frac{p_5}{5p_5} = \frac{6}{5}\leq \beta.
\end{align*}
	If $J_1$ is processed on the same machine as one job other than $J_6$ and $J_7$ in $\sigma^*$, then  $C^* \geq p_1+p_5 \geq C^L$ by (\ref{e31}).  If $J_1$ is processed on the same machine as both $J_6$ and $J_7$ in $\sigma^*$, then $C^* \geq p_1+p_6+p_7\geq p_1+2p_7$. Therefore, by (\ref{e31}),
	\[
	\frac{C^L}{C^*} 
	\leq \frac{\frac{2}{5}p_1+ \frac{3}{5}(p_2+p_7)+p_5}{\max\{p_1+2p_7, p_2+p_5\}}\leq \frac{\frac{2}{5}(p_1+2p_7) + \frac{4}{5}(p_2+p_5)}{\max\{p_1+2p_7, p_2+p_5\}} \leq \frac{6}{5}\leq \beta.
	\]
	Therefore, we assume in the following that $J_1$ is processed on the same machine as  one of $J_6$ and $J_7$  in $\sigma^*$ and thus  $C^* \geq p_1+p_7$. 
	
	If $J_2$ is processed on the same machine as at least two other jobs in $\sigma^*$, then $C^* \geq p_2+p_5+p_7 \geq C^L$ by (\ref{e31}). If $J_2$ is processed on the same machine as $J_6$ or $J_7$ in $\sigma^*$, then $C^* \geq p_3+p_4+p_5 \geq C^L$ by (\ref{e31}). Therefore, we only need to consider the case where $J_2$ is processed on the same machine as one of $J_3$, $J_4$ and $J_5$ in $\sigma^*$. Thus $C^* \geq p_2+p_5$.
	
According to the above discussion regarding the assignment of $J_1$ and $J_2$ in $\sigma^*$, it follows that the machine that does not process $J_1$ and $J_2$ processes $3$ jobs and at least two of them have a processing time of at least $p_5$. Thus $C^* \geq 2p_5+p_7$. Therefore, by (\ref{e31}),
\begin{align*}	
	\frac{C^L}{C^*} &
	\leq \frac{\frac{2}{5}p_1 + \frac{3}{5}(p_2+p_7) +p_5}{\max\{p_1+p_7, p_2+p_5, 2p_5+p_7\}} = \frac{\frac{2}{5}(p_1+p_7) + \frac{3}{5}(p_2+p_5) + \frac{1}{5}(2p_5+p_7)}{\max\{p_1+p_7, p_2+p_5, 2p_5+p_7\}} \leq \frac{6}{5}\leq \beta.
\end{align*}

If $p_3<\alpha p_7$, the partial schedule of $\sigma^L$ is one of $\sigma^{(k)}$, where $k=4,5,6$. If $J_5$ determines the makespan of $\sigma^L$, then $C^L \leq  p_3+p_5 \leq C^*$ by Lemma \ref{tho}(i). If $J_4$ determines the makespan of $\sigma^L$, since $J_4$ is assigned to the machine where it can be started earliest, then
\begin{equation}\label{e32}
	C^L = \min\{p_1,p_2+p_7, p_3+p_5\} + p_4.
\end{equation}
If $\mathcal{J}_1^* = \{J_1\}$, then by Lemma \ref{tho}(ii) and \eqref{t72}, \eqref{t71}, we have
\[
\frac{C^L}{C^*} \leq \frac{p_3+p_4+p_5}{p_3+p_5+p_7} \leq 1 + \frac{p_4-p_7}{p_3+p_5+p_7} \leq 1 + \frac{p_3-p_7}{p_3+2p_7} \leq 1 + \frac{\alpha-1}{\alpha+2} = \frac{2\alpha+1}{\alpha+2} = \beta.
\]
If $J_1$ is processed on the same machine as one job of $J_2$, $J_3$ and $J_4$ in $\sigma^*$, then  $C^* \geq p_1+p_4 \geq C^L$ by (\ref{e32}). Therefore, we assume in the following that $J_1$ is processed on the same machine as at least one of $J_5$, $J_6$ and $J_7$ in $\sigma^*$. Then in $\sigma^*$, the machine processing at least three jobs must process a job of processing time at least $p_4$. Thus, $C^* \geq p_1+p_7$ and $C^* \geq p_4+2p_7$. Therefore, by (\ref{e32}), \eqref{t72} and \eqref{t71}, we have
\begin{align*}	
	\frac{C^L}{C^*} &\leq \frac{p_1+p_4}{\max\{p_1+p_7, p_4+2p_7\}} = \frac{p_1+p_7+p_4-p_7}{\max\{p_1+p_7, p_4+2p_7\}} \leq 1 + \frac{p_4-p_7}{p_4+2p_7} \\
	&\leq 1 + \frac{\alpha p_7-p_7}{\alpha p_7+2p_7} = 1 + \frac{\alpha-1}{\alpha+2}=\frac{2\alpha+1}{\alpha+2} = \beta.
\end{align*}
\end{proof}

By Lemmas \ref{thr2} and \ref{thr3}, we have the following theorem. 
\begin{theorem}\label{thm1}
The competitive ratio of DLPT is $\frac{13+\sqrt{15}}{14}$ for $m=3$ and $\tau=0$.   
\end{theorem}

\section{Lower Bounds}

In this section, we present several lower bounds for the problem. Since all jobs are available at time $0$, any reasonable algorithm proceeds without introducing idle time on any machine. Specifically, at time $0$, $m$ jobs start processing on different machines. Let  $\mathcal{J}_{0}^A$ denote the set of $m$ jobs selected by Algorithm A to start processing at time  $0$.

\begin{theorem}\label{lb1}
No online algorithm for $\tau=0$ can have a competitive ratio less than $\frac{7}{6}$ and $\beta$ for $m=2$ and $m\geq 3$, respectively. 
\end{theorem}
\begin{proof}
The proof of the lower bound for $\tau=0$ and $m=2$ is presented in Table \ref{lbi2}. We enumerate all possible schedules for instances containing $5$ jobs. In most cases, considering the assignment of the first few jobs is sufficient. For example, as long as $J_1\notin\mathcal{J}_{0}^A$, then $\frac{C^A(I_1)}{C^*(I_1)}\geq \frac{5}{4}$.
There are also some scenarios where it is necessary to consider the assignment of more jobs. For instance, if $\mathcal{J}_{0}^A=\{J_1, J_4\}$, it is necessary to further consider which job is the third one to be processed to establish that the competitive ratio is no less than $\frac{7}{6}$. At this point, $J_4$ has been completed, so its processing time is already determined. Therefore, in instances $I_2$, $I_4$ and $I_5$, the processing time of $J_4$ must be identical. However, $J_1$ is still in process when the third job is determined, so its processing times may differ across different instances. For the scenario corresponding to instance $I_3$, the third started job can be either $J_4$ or $J_5$. Therefore, the processing times of $J_4$ and  $J_5$ in $I_3$ are also the same.

The proof for $\tau=0$ and $m=3$ is similar, and we present it in Table \ref{lbi3}. In this case, we enumerate all possible schedules for instances containing $7$ jobs, and need to discuss at most up to the fifth started job. Similarly, if multiple instances are used in scenarios where the first few processed jobs are the same but the subsequent ones differ, then the processing times of the completed jobs must be equal across these instances. 

Next, we generalize the proof from the case of $m=3$ to $m > 3$. This is achieved by introducing $m-3$ dummy jobs with sufficiently large processing times to the instance constructed for $m=3$. For clarity, we index the jobs as $ J_{-(m-4)}, J_{-(m-3)}, \dots, J_{-1}, J_0, J_1, \dots, J_7 $, where $ J_j $, $ j \leq 0 $, denotes a dummy job. For the non-dummy jobs, we retain the instance from Table \ref{lbi3} and define the processing times of the dummy jobs as the optimal makespan values of the corresponding instances.
If at least one of the dummy jobs does not belong to $\mathcal{J}_{0}^A$, then the instance corresponding to $I_1$ demonstrates that the ratio of the makespan of the schedule produced by the algorithm to that of the optimal schedule is at least $\frac{4}{3}$. If all dummy jobs belong to $\mathcal{J}_{0}^A$, the instance in Table \ref{lbi3} corresponding to the remaining jobs in $\mathcal{J}_{0}^A$ should be selected.  Note that the processing times of the dummy jobs are no less than those of the non-dummy jobs, and their completion times are no less than the start times of the jobs determining the makespan. Consequently, both the optimal makespan and the makespan produced by Algorithm A remain unchanged. Thus, the lower bound established for $m=3$ holds for any $m > 3$.
\end{proof}

\begin{table}
\begin{footnotesize}
	\begin{tabular}{|c|c|c|c|c|c|c|c|c|c|}
	\hline
    1,2-SJ & \multicolumn{2}{c|}{1-CJ\&P} & 3-SJ &  Instance & $\sigma^A$ & $C^A$ & $\sigma^*$ & $C^*$ & $\frac{C^A}{C^*}$ \\
	\hline
	$J_a, J_b$, $a,b>1$ & & & &  $I_1: \{4, 1, 1, 1, 1\}$ & $\{4, 1\}, \{1, 1, 1\}$ & $5$ & $\{4\}, \{1, 1, 1, 1\}$ & $4$ & $\frac{5}{4}$ \\
	\hline
	$J_1, J_2$ & & & &  $I_2: \{3, 3, 2, 2, 2\}$ & $\{3, 2, 2\}, \{3, 2\}$ & $7$ & $\{3, 3\}, \{2, 2, 2\}$ & $6$ & $\frac{7}{6}$ \\
	\hline
	\multirow{2}{*}{$J_1, J_3$} & \multirow{2}{*}{$J_3$} & \multirow{2}{*}{2} & $J_2$ &  $I_2: \{3, 3, 2, 2, 2\}$ & $\{3, 2, 2\}, \{3, 2\}$ & $7$ & $\{3, 3\}, \{2, 2, 2\}$ & $6$ & $\frac{7}{6}$ \\
	\cline{4-10}
	 &   &   & $J_a$, $a\in\{4,5\}$  &  $I_3: \{3, 3, 2, 1, 1\}$ & $\{3, 3\}, \{2, 1, 1\}$ & $6$ & $\{3, 2\}, \{3, 1, 1\}$ & $5$ & $\frac{6}{5}$ \\
	\hline  
	 &  &  & $J_2$ &  $I_2: \{3, 3, 2, 2, 2\}$ & $\{3, 2, 2\}, \{3, 2\}$ & $7$ & $\{3, 3\}, \{2, 2, 2\}$ & $6$ & $\frac{7}{6}$ \\
	\cline{4-10}
	$J_1, J_4$ & $J_4$ & $2$ & $J_3$ &  $I_4: \{4, 4, 2, 2, 0\}$ & $\{4, 4\}, \{2, 2, 0\}$ & $8$ & $\{4, 2\}, \{4, 2, 0\}$ & $6$ & $\frac{4}{3}$ \\
	\cline{4-10}
	&  &  & $J_5$ &  $I_5: \{5, 2, 2, 2, 1\}$ & $\{5, 2\}, \{2, 2, 1\}$ & $7$ & $\{5, 1\}, \{2, 2, 2\}$ & $6$ & $\frac{7}{6}$ \\
	\hline
	$J_1, J_5$ &  &  &   &  $I_5: \{5, 2, 2, 2, 1\}$ & $\{5, 2\}, \{2, 2, 1\}$ & $7$ & $\{5, 1\}, \{2, 2, 2\}$ & $6$ & $\frac{7}{6}$ \\
	\hline
\end{tabular}
\end{footnotesize}
		\caption{Proof of the lower bound for $m=2$ and $\tau=0$. \\
	(1,2-SJ=The first two started jobs, 3-SJ=The third started job, 1-CJ\&P=The first completed job and its processing time.) 
}\label{lbi2}
\end{table}

\begin{table}
		\begin{tabular}{|c|c|c|c|c|c|c|c|c|}
	\hline
	\multicolumn{1}{|c|}{1,2,3-SJ} & \multicolumn{2}{c|}{1-CJ\&P}   & \multicolumn{1}{c|}{4-SJ} & \multicolumn{2}{c|}{2-CJ\&P} & \multicolumn{1}{c|}{5-SJ} & \multicolumn{1}{c|}{No.} \\
	\hline
	$J_a, J_b, J_c$, $a,b,c>1$ &   &   &  & & & &  $I_1$ \\
	\hline
	 &  &  & $J_4$ &   & & &  $I_2$ \\
\cline{4-8}
	$J_1, J_2, J_3$ & $J_3$  & $12$  & $J_a$, $a\in\{5,6\}$ &  & & & $I_3$  \\
\cline{4-8}
	&   &   & $J_7$ &  & & & $I_4$  \\
	\hline
	\multirow{4}{*}{$J_1, J_2, J_4$} & \multirow{4}{*}{$J_4$} & \multirow{4}{*}{$12$}  & $J_3$ &  & & & $I_2$  \\
\cline{4-8}
	 &   &  & $J_a$, $a\in\{5,6\}$ &  & & & $I_3$  \\
\cline{4-8}
	&   &   & \multirow{2}{*}{$J_7$} &  \multirow{2}{*}{$J_7$} & \multirow{2}{*}{$4$} & $J_3$ & $I_5$  \\
\cline{7-8}
	&   &   &  & & & $J_a$, $a\in\{5,6\}$ &  $I_6$ \\
	\hline
	$J_1, J_2, J_a$, $a\in\{5,6\}$ &   & &  &  & & & $I_7$  \\
	\hline
	\multirow{4}{*}{$J_1, J_2, J_7$} & \multirow{4}{*}{$J_7$}  & \multirow{4}{*}{$4$}  & $J_3$ &   & &  &  $I_5$ \\
\cline{4-8}
	&   &   & \multirow{2}{*}{$J_4$} & \multirow{2}{*}{$J_4$} & \multirow{2}{*}{$12$} & $J_3$ & $I_5$  \\
\cline{7-8}
	&   &   &   & &  & $J_a$, $a\in\{5,6\}$ & $I_6$  \\
\cline{4-8}
	&   &   & $J_a$, $a\in\{5,6\}$ &  & &  & $I_8$  \\
	\hline
	$J_1, J_a, J_b$, $a,b\in\{3,4,5,6\}$ &   &  & &  & & & $I_{9}$  \\
	\hline
	\multirow{3}{*}{$J_1, J_a, J_7$, $a\in\{3,4\}$} & \multirow{3}{*}{$J_7$} & \multirow{3}{*}{$4$} & $J_b$, $b\in\{3,4,5,6\}\setminus \{a\}$ & & &   & $I_{10}$  \\
\cline{4-8}
&   &   & \multirow{2}{*}{$J_2$} & \multirow{2}{*}{$J_a$} & \multirow{2}{*}{$4\alpha$} & $J_c$, $c\in\{3,4\}\setminus \{a\}$  & $I_{11}$  \\
\cline{7-8}
&   &   &  & & & $J_c$, $c\in\{5,6\}$  &  $I_{12}$ \\
\hline
	\multirow{2}{*}{$J_1, J_5, J_7$} & \multirow{2}{*}{$J_7$} & \multirow{2}{*}{$0$} & $J_2$ &  & & & $I_{13}$  \\
\cline{4-8}
&   &   & $J_a$, $a\in\{3,4,6\}$ &   & & & $I_{9}$  \\
\hline
	$J_1, J_6, J_7$ &   & &  &  & & &  $I_{14}$ \\
	\hline
\end{tabular}

	\vspace{0.5cm} 	

\begin{footnotesize}
	\begin{tabular}{|c|c|c|c|c|c|c|}
		\hline
		\multicolumn{1}{|c|}{No.} & \multicolumn{1}{c|}{Instance} & $\sigma^A$ & $C^A$ & $\sigma^*$ & $C^*$ & $\frac{C^A}{C^*}$ \\
		\hline
		$I_1$ & $\{12, 4, 4, 4, 4, 4, 4\}$ & $\{12, 4\}, \{4, 4, 4\}, \{4, 4\}$ & $16$ & $\{12\}, \{4, 4, 4\}, \{4, 4, 4\}$ & $12$ & $\frac{4}{3}$ \\
		\hline
		$I_2$ & $\{15, 15, 12, 12, 9, 9, 9\}$ & $\{15, 9, 9\}, \{15, 9\}, \{12, 12\}$ & $33$ & $\{15, 12\}, \{15, 12\}, \{9, 9, 9\}$ & $27$ & $\frac{11}{9}$ \\
		\hline
		$I_3$ & $\{18, 18, 12, 12, 6, 6, 0\}$ & $\{18, 12\}, \{18, 6\}, \{12, 6,0\}$ & $30$ & $\{18, 6\}, \{18,6\}, \{12, 12, 0\}$ & $24$ & $\frac{5}{4}$ \\
		\hline
		$I_4$ & $\{24, 15, 12, 9, 9, 9, 3\}$ & $\{24, 9\}, \{15, 9\}, \{12, 9, 3\}$ & $33$ & $\{24, 3\}, \{15, 12\}, \{9, 9, 9\}$ & $27$ & $\frac{11}{9}$ \\
		\hline
		$I_5$ & $\{32, 20, 16, 12, 12, 12, 4\}$ & $\{32, 12\}, \{20, 12\}, \{16, 12, 4\}$ & $44$ & $\{32, 4\}, \{20, 16\}, \{12, 12, 12\}$ & $36$ & $\frac{11}{9}$ \\
		\hline
		$I_6$ & $\{28, 28, 28, 12, 12, 12, 4\}$ & $\{28, 28\}, \{28, 12\}, \{12, 12, 4\}$ & $56$ & $\{28, 12, 4\}, \{28, 12\}, \{28, 12\}$ & $44$ & $\frac{14}{11}$ \\
		\hline
		$I_7$ & $\{12, 12, 8, 8, 4, 4, 0\}$ & $\{12, 8\}, \{12, 4\}, \{8, 4, 0\}$ & $20$ & $\{12, 4\}, \{12, 4\}, \{8, 8, 0\}$ & $16$ & $\frac{5}{4}$ \\
		\hline
		$I_8$ & $\{132, 132, 88, 88, 44, 44, 4\}$ & $\{132, 88\}, \{132, 44\}, \{88, 44, 4\}$ & $220$ & $\{132, 44\}, \{132, 44\}, \{88, 88, 4\}$ & $180$ & $\frac{11}{9}$ \\
		\hline
		$I_{9}$ & $\{16, 16, 4, 4, 4, 4, 0\}$ & $\{16, 4\}, \{16, 4\}, \{4, 4, 0\}$ & $20$ & $\{16\}, \{16\}, \{4, 4, 4, 4, 0\}$ & $16$ & $\frac{5}{4}$ \\
		\hline
$I_{10}$ & $\{176, 176, 44, 44, 44, 44, 4\}$ & $\{176\}, \{176, 44\}, \{44, 44, 44, 4\}$ & $220$ & $\{176\}, \{176\}, \{44, 44, 44, 44, 4\}$ & $180$ & $\frac{11}{9}$ \\
\hline
$I_{11}$ & \makecell{$\{8\alpha-4, 8\alpha-4, 4\alpha, 4\alpha,$ \\ $6\alpha-4, 6\alpha-4, 4\}$}  & \makecell{$\{8\alpha-4, 6\alpha-4\},  $\\ $\{4\alpha, 4\alpha, 6\alpha-4\}, \{8\alpha-4, 4\}$} & \scriptsize $14\alpha-4$  & \makecell{$\{8\alpha-4, 4\alpha\}, \{8\alpha-4, 4\alpha\},$\\ $\{6\alpha-4, 6\alpha-4, 4\}$} & \scriptsize $12\alpha-4$ &  $\beta$ \\
\hline
$I_{12}$ & $\{4\alpha+4, 4\alpha, 4\alpha, 4\alpha, 4, 4, 4\}$ & $\{4\alpha+4, 4\alpha\}, \{4\alpha, 4, 4\}, \{4\alpha, 4\}$ & \scriptsize $8\alpha+4$ & $\{4\alpha+4, 4\}, \{4\alpha, 4, 4\}, \{4\alpha, 4\alpha\}$ & \scriptsize $4\alpha+8$ & $\beta$ \\
		\hline
		$I_{13}$ & $\{12, 12, 8, 8, 4, 0, 0\}$ & $\{12, 8\}, \{8, 4\}, \{12, 0, 0\}$ & $20$ & $\{12, 4\}, \{12\}, \{8, 8, 0, 0\}$ & $16$ & $\frac{5}{4}$ \\
\hline
		$I_{14}$ & $\{8, 8, 8, 8, 8, 4, 4\}$ & $\{8, 8\}, \{8, 8, 4\}, \{8, 4\}$ & $20$ & $\{8, 8\}, \{8, 8\}, \{8, 4, 4\}$ & $16$ & $\frac{5}{4}$ \\
		\hline
	\end{tabular}
\end{footnotesize}

		\caption{Proof of the lower bound for $m=3$ and $\tau=0$. \\
	(1,2,3-SJ=The first three started jobs, 4-SJ=The forth started job, 5-SJ=The fifth started job, 1-CJ\&P=The first completed job and its processing time, 2-CJ\&P=The second completed job and its processing time, No.=The number of instance.) 
}
\label{lbi3}
\end{table}

\begin{theorem}\label{lb2}
	No online algorithm can have a competitive ratio less than $\frac{5}{4}$ for $m\geq 2$ and $\tau\geq 1$. 
\end{theorem}
\begin{proof}
Consider instances with $m+2$ jobs. 
If $\{J_{\pi(m+1)}, J_{\pi(m+2)}\} \subseteq \mathcal{J}^A_0$, let the processing times of the jobs be 
\[
p_{1} = \dots = p_{m-1} = 3, \quad p_{m} = p_{m+1} = p_{m+2} = 1,
\]
and set the permutation to be $\pi(i)=i$ for $1\leq i\leq m+2$. 
Obviously, $\tau(\pi)=0$. Regardless of the processing order of the remaining jobs, we have $C^A \ge 4$ and $C^* = 3$. Thus, $\frac{C^A}{C^*} \geq \frac{4}{3}$.

If $\{J_{\pi(m+1)}, J_{\pi(m+2)}\} \nsubseteq \mathcal{J}^A_0$, let the processing times of the jobs be
\[
p_1 = \dots = p_{m-1} = 3, \quad p_m = p_{m+1} = 2, \quad p_{m+2} = 1.
\]
Obviously, $C^* = 4$. Let $\pi_1 = (1, 2, \dots, m+2)$ and $\pi_2 = (1, 2, \dots, m, m+2, m+1)$ be two permutations satisfying $\tau(\pi_1)=0$ and $\tau(\pi_2)=1$. If $\{J_{\pi(m)}, J_{\pi(m+1)}\} \subseteq \mathcal{J}^A_0$, we set the permutation to $\pi_1$. If $\{J_{\pi(m)}, J_{\pi(m+2)}\} \subseteq \mathcal{J}^A_0$, we set the permutation to $\pi_2$. In either case, the algorithm starts processing two jobs with processing time $2$ at time $0$. Consequently, $C^A \ge 5$ and $\frac{C^A}{C^*}\geq \frac{5}{4}$.

If exactly one of $J_{\pi(m)}, J_{\pi(m+1)}$, and $J_{\pi(m+2)}$ belongs to $\mathcal{J}^A_0$, then regardless of whether the permutation is $\pi_1$ or $\pi_2$, that job is the earliest finished job in $\mathcal{J}^A_0$. First, assume that $J_{\pi(m)} \in \mathcal{J}^A_0$. If the $(m+1)$-th started job is $J_{\pi(m+1)}$, we set the permutation to $\pi_2$. Otherwise, we set the permutation to $\pi_1$. The processing time of this job is $1$, and the start time of the remaining job of processing time $2$ is no earlier than $3$. Thus $C^A \geq 5$ and $\frac{C^A}{C^*}\geq \frac{5}{4}$. Next, assume that $J_{\pi(m)} \notin \mathcal{J}^A_0$. We set the permutation to $\pi_2$ if $J_{\pi(m+1)} \in \mathcal{J}^A_0$, and $\pi_1$ if $J_{\pi(m+2)} \in \mathcal{J}^A_0$. In both cases, the job with the earliest completion time in $\mathcal{J}^A_0$ has a processing time of $1$, and the next job processed on the same machine has a processing time of $2$. The start time of the remaining job of processing time $2$ is no earlier than $3$.  Thus, $C^A \geq 5$ and $\frac{C^A}{C^*}\geq \frac{5}{4}$.
\end{proof}	

\begin{theorem}\label{lb3}
	For any $0\leq t\leq m-2$, no online algorithm can have a competitive ratio less than $2 - \frac{1}{m - t}$ for $m\geq 2$ and $\overline{\tau}\geq m^2-(m+1)t-m$. 
\end{theorem}
\begin{proof}
For any $0 \leq t \leq m - 2$, consider an instance with $n = 2m - t - 1$ jobs, including $t+1$ {\it big} jobs with processing time $m - t$, $m - t - 1$ {\it medium} jobs with processing time $m - t - 1$, and $m-t-1$ {\it small} jobs with processing time $1$. Obviously, $C^* = m - t$. We show that for any algorithm $A$, there exists a permutation $\pi$ with $\overline{\tau}(\pi)\leq m^2 - (m + 1)t - m$ such that if Algorithm $A$
assigns jobs according to the order given by $\pi$, then $C^A \geq 2m - 2t - 1$. Thus $\frac{C^A}{C^*} \geq \frac{2m - 2t - 1}{m - t} = 2 - \frac{1}{m - t}$, which completes the proof.

Assume that $\mathcal{J}^A_0=\{J_{\pi(j_1)},\dots, J_{\pi(j_m)}\}$, where $j_1<j_2<\dots<j_m$. We construct the permutation $\pi$ such that $\mathcal{J}^A_0$ contains exactly $t$ big jobs, $m - t - 1$ medium jobs, and one small job. In more detail,
\begin{equation}\label{t81}
\pi(j_i) = 
\begin{cases} 
	i & 1 \le i \le t \\ 
	i+1 & t+1 \le i \le m-1 \\ 
	2m-t-1 & i=m 
\end{cases}
\end{equation}
Among these jobs, the small job must complete the earliest. 
The permutation $\pi$ is constructed such that the subsequent $m - t - 2$ jobs selected by algorithm $A$, $J_{\pi(j_{m+1})},\dots, J_{\pi(j_{2m-t-2})}$, are all small jobs, i.e., 
\begin{equation}\label{t82}
	\pi(j_i) = i, \quad m+1 \le i \le 2m-t-2.
\end{equation} 
Once all these jobs have been completed, the machine with the minimum load has a load of $m-t-1$. 
Finally, $\pi(j_{2m-t-1}) = t+1$ and the last job is a big job.  
Consequently, $C^A \geq 2m - 2t - 1$.

Next, we estimate the number of true inversions in $\pi$ (Ref. Figure \ref{fig4}). 
Clearly, for any $1 \le i < l \le m$, $j_i < j_l$ and $\pi(j_i) < \pi(j_l)$. Thus $(j_i,j_l)$ is not an inversion. 
For any $j_i$, $m+1 \le i \le 2m-t-2$, $(j_i, j_l)$ may be a true inversion only if $J_{j_l}$ is a big or medium job. 
Apart from these true inversions, $(j_i, j_{2m-t-1})$ may be a true inversion only if $J_{j_i}$ is a medium job or $J_{j_i}$ is the small job $J_{j_m}$. 
Therefore,
\[
\overline{\tau}(\pi) \le (t+1+m-t-1)(m-t-2) + m-t-1+1 = m^2 - (m+1)t - m.
\]

\begin{figure}
	\includegraphics[width=140mm]{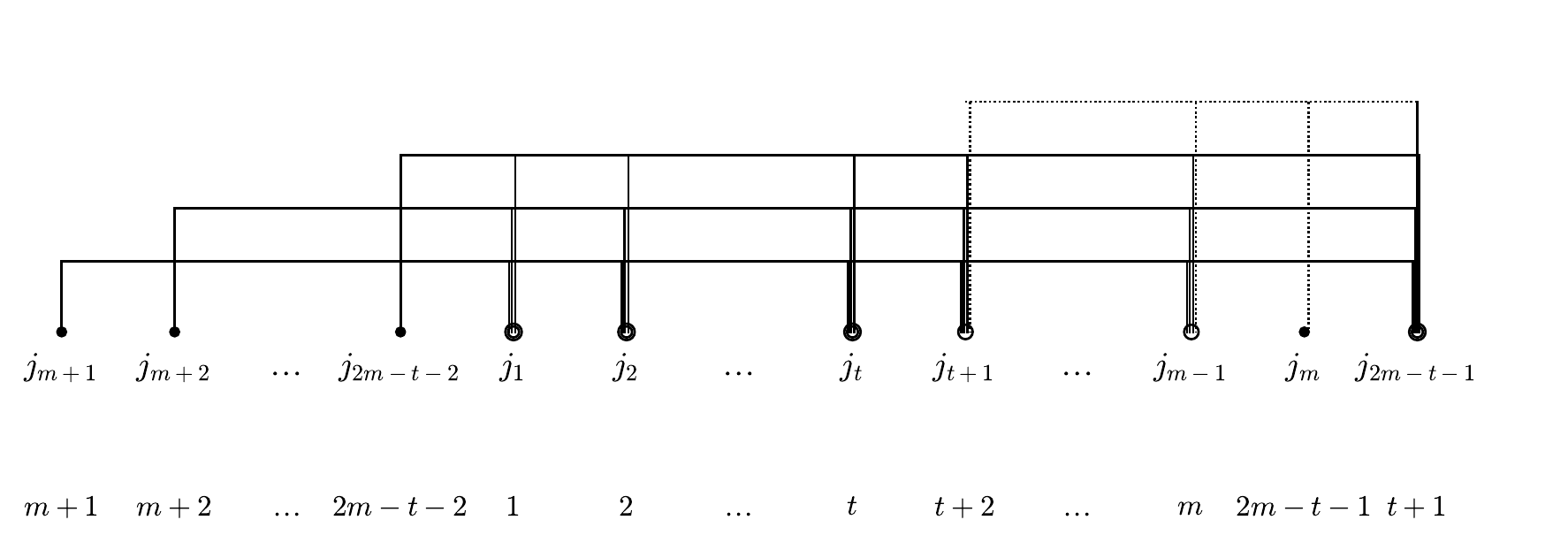}
	\centering
	\caption{True inversions in $\pi$. }\label{fig4}
\end{figure}

From the above discussion and Lemma \ref{ln2}, it follows that if the upper bound for the number of inversions in the predicted order is at least $m^2-(m+1)t-m$, the competitive ratio of any algorithm is at least $2 - \frac{1}{m - t}$.
\end{proof}	

By Theorems \ref{thm0} and \ref{lb1}, \ref{lb2} \ref{lb3}, we know that Algorithm POLPT is optimal for any $\tau \geq 0$ when $m=2$, for $1\leq \tau \leq 3$ and $\tau\geq 6$ when $m=3$, and for $2\leq \tau \leq 3$ and $\tau \geq m^2-m$ when $m\geq 4$. By Theorems \ref{thm1} and \ref{lb1}, we know that Algorithm DLPT is optimal for $\tau=0$ and $m=3$. 

\section{Conclusions}

This paper investigates the non-clairvoyant online parallel machine scheduling problem where the order of job processing times is predictable. The actual processing time of a job remains unknown until its completion, but its relative position among the processing times of all jobs can be predicted in advance, although this predicted information may contain errors. We use the number of inversions between the predicted order and the true order as a metric to measure prediction accuracy and perform a competitive analysis using this metric as a parameter. We derive lower bounds for the problem, design algorithms, and prove their competitive ratios. Optimal algorithms are achieved when the number of machines is small and prediction errors are either small or large.

There are many problems worthy of further research. First and foremost, it is crucial to determine optimal algorithms for cases that have not yet been solved, particularly the two seemingly simplest cases. The case $\tau=0$ is equivalent to adding the condition that the order of job processing times is known. While optimal algorithms are known for arbitrary $m$ without this condition, the addition of this condition appears to significantly increase the difficulty. Insights from $m=3$ imply that a case-by-case analysis may be unavoidable to derive the optimal algorithms for any $m\geq 4$. It is unknown whether obtaining the optimal algorithm for sufficiently large $m$ would be easier. For $m=3$, optimal algorithms are lacking only for $\tau=4$ and $\tau=5$, intermediate states that seem difficult to handle. Secondly, other definitions measuring the difference between two permutations can be used to define prediction error. Additionally, scenarios where predictions for both order and magnitude of processing times exist simultaneously can be explored. Finally, corresponding problems under other machine environments and with different objective functions can also be investigated.

\end{document}